%% file: manuscript.tex
\documentclass[manuscript]{acmart}

\usepackage{multirow}
\usepackage{xcolor}

\usepackage{pifont}
\newcommand{\cmark}{\ding{51}}
\newcommand{\xmark}{\ding{55}}

\usepackage{booktabs}
\usepackage{threeparttable}
\usepackage{adjustbox}
\usepackage{xspace}

\newcommand{\athena}{\mbox{\textsc{A-THENA}}\xspace}

\newcommand{\ciciot}{\mbox{\textsc{CICIoT23-WEB}}\xspace}
\newcommand{\mqtt}{\mbox{\textsc{MQTT-IoT-IDS2020}}\xspace}
\newcommand{\iotid}{\mbox{\textsc{IoTID20}}\xspace}

\AtBeginDocument{%
  }

\setcopyright{acmlicensed}
\copyrightyear{2026}
\acmYear{2026}
\acmDOI{10.1145/3811033}

\makeatletter
\def\acmJournal#1{} 
\def\acmVolume#1{}
\def\acmNumber#1{}
\def\acmArticle#1{}
\def\acmMonth#1{}
\makeatother

\acmJournal{TAISAP} 
\acmVolume{}
\acmNumber{}
\acmArticle{}
\acmMonth{}

\begin{document}

\title[A-THENA]{A-THENA: Early Intrusion Detection for IoT with Time-Aware Hybrid Encoding and Network-Specific Augmentation}

\author{Ioannis Panopoulos}
\email{ioannispanop@mail.ntua.gr}
\orcid{0009-0005-5364-4410}
\affiliation{%
  \institution{National Technical University of Athens}
  \city{Athens}
  \country{Greece}
}

\author{Maria Lamprini A. Bartsioka}
\email{bartsiokamarilina@mail.ntua.gr}
\orcid{0009-0007-4238-1373}
\affiliation{%
  \institution{National Technical University of Athens}
  \city{Athens}
  \country{Greece}
}

\author{Sokratis Nikolaidis}
\email{sokratisnikolaidis@mail.ntua.gr}
\orcid{0009-0004-3439-7402}
\affiliation{%
  \institution{National Technical University of Athens}
  \city{Athens}
  \country{Greece}
}

\author{Stylianos I. Venieris}
\email{s.venieris@samsung.com}
\orcid{0000-0001-5181-6251}
\affiliation{%
  \institution{Samsung AI Center}
  \city{Cambridge}
  \country{UK}
}

\author{Dimitra I. Kaklamani}
\email{dkaklam@mail.ntua.gr}
\orcid{0000-0002-1641-9485}
\affiliation{%
  \institution{National Technical University of Athens}
  \city{Athens}
  \country{Greece}
}

\author{Iakovos S. Venieris}
\email{venieris@cs.ece.ntua.gr}
\orcid{0000-0003-3011-3746}
\affiliation{%
  \institution{National Technical University of Athens}
  \city{Athens}
  \country{Greece}
}

\renewcommand{\shortauthors}{I. Panopoulos et al.}

\begin{abstract}
  The proliferation of Internet of Things (IoT) devices has significantly expanded attack surfaces, making IoT ecosystems particularly susceptible to sophisticated cyber threats. To address this challenge, this work introduces \athena, a lightweight early intrusion detection system (EIDS) that significantly extends preliminary findings on time-aware encodings. \athena employs an advanced Transformer-based architecture augmented with a generalized Time-Aware Hybrid Encoding (THE), integrating packet timestamps to effectively capture temporal dynamics essential for accurate and early threat detection. The proposed system further employs a Network-Specific Augmentation (NA) pipeline, which enhances model robustness and generalization. We evaluate \athena on three benchmark IoT intrusion detection datasets---\ciciot, \mqtt, and \iotid---where it consistently achieves strong performance. Averaged across all three datasets, it improves accuracy by 6.88 percentage points over the best-performing traditional positional encoding, 3.69 points over the strongest feature-based model, 6.17 points over the leading time-aware alternatives, and 5.11 points over related models, while achieving near-zero false alarms and false negatives. To assess real-world feasibility, we deploy \athena on the Raspberry Pi Zero 2 W, demonstrating its ability to perform real-time intrusion detection with minimal latency and memory usage. These results establish \athena as an agile, practical, and highly effective solution for securing IoT networks.
\end{abstract}



\received{5 August 2025}
\received[revised]{2 April 2026}
\received[accepted]{9 April 2026}

\maketitle

\section{INTRODUCTION}
\label{sec:intro}
\input{sections/1_intro}

\section{BACKGROUND AND RELATED WORK}
\label{sec:back_related}
\input{sections/2_back_related}

\section{OUR APPROACH}
\label{sec:system}
\input{sections/3_approach}

\section{IMPLEMENTATION}
\label{sec:implementation}
\input{sections/4_implementation}

\section{EXPERIMENTAL METHODOLOGY}
\label{sec:exp_meth}
\input{sections/5_exp_methodology}

\section{RESULTS}
\label{sec:results}
\input{sections/6_results}

\section{CONCLUSION}
\label{sec:conclusion}
\input{sections/7_conclusion}

\bibliographystyle{ACM-Reference-Format}
\bibliography{references}


\end{document}

%% file: sections/1_intro.tex
The Internet of Things (IoT) has transformed digital environments by enabling smart devices to autonomously sense, process, and exchange data in real time across diverse domains such as healthcare, smart homes, and industrial automation~\cite{iot22real}. These devices, built upon embedded systems with networking capabilities, operate under strict resource constraints, including limited processing power, memory, and energy availability. As a result, there is a critical need for compact hardware architectures, lightweight software, and efficient data-processing models to support real-time, reliable operation at the network edge.

The widespread adoption of IoT devices has introduced significant security vulnerabilities, largely due to their limited computational resources, weak authentication mechanisms, and frequent exposure to untrusted networks. These vulnerabilities make IoT ecosystems prime targets for cyberattacks, ranging from malware infections and Distributed Denial-of-Service (DDoS) attacks to more sophisticated Man-in-the-Middle (MitM) exploits and adversarial manipulation of data streams. Ensuring the security of IoT systems is therefore critical~\cite{omolara22compsec}, as security breaches can lead to data leaks, service disruptions, and potential physical harm in safety-critical applications. More broadly, the long-term success of the IoT paradigm hinges on user trust. Consumers, industries, and institutions are unlikely to adopt and integrate IoT technologies at scale unless they are confident in the security and trustworthiness of the underlying infrastructure.

To mitigate such security risks, Intrusion Detection Systems (IDS) have been widely adopted. IDS solutions monitor network traffic or host activity to detect malicious behavior and potential intrusions. In recent years, Machine Learning (ML) and, more prominently, Deep Learning (DL), have become integral to the development of IDS solutions, driven by their ability to detect complex attack patterns that traditional rule-based systems fail to recognize. Unlike conventional IDS, which rely on manually crafted signatures or predefined heuristics, ML-based IDS can learn from data, enabling the detection of previously unseen threats and zero-day attacks. Particularly in the context of IoT security, where network traffic is highly dynamic and attacks can evolve rapidly, DL models such as Convolutional Neural Networks (CNNs)~\cite{mohammadpour22apllied}, Recurrent Neural Networks (RNNs)~\cite{ullah22acccess}, Transformers~\cite{manocchio24expert}, and hybrid architectures~\cite{saiyedand24tmlcn} have been increasingly employed to analyze network flows, uncover anomalies, and improve detection accuracy. These models excel at processing large-scale traffic data, automatically extracting meaningful representations, and generalizing beyond known attack signatures. The shift towards ML-driven IDS marks a significant advancement in cybersecurity, providing adaptive, scalable, and real-time threat detection capabilities essential for protecting modern IoT ecosystems.

A key challenge in intrusion detection is the need for real-time recognition to minimize response times and mitigate the impact of attacks. This has led to the emergence of Early Intrusion Detection Systems (EIDS)~\cite{lopez19nca}, which aim to classify and detect intrusions as early as possible within a network session. Recent DL research has demonstrated the strong performance of Transformers for sequential data processing, particularly in NLP~\cite{patwardhan23info}. However, unlike text, network traffic flows are time series, where packet arrival times convey critical contextual information often overlooked by existing IDS models. To address this limitation, this paper proposes \athena, a complete Transformer-based EIDS. By significantly extending our initial exploration of time-aware encodings~\cite{splitech_paper}, \athena effectively models both sequence structure and temporal dynamics, resulting in superior intrusion detection capabilities. Grounded in these foundational concepts, this study makes the following primary contributions:
\begin{itemize}
    \item A complete, lightweight end-to-end early intrusion detection system, named \athena, which introduces the Time-Aware Hybrid Encoding (THE) mechanism; an adaptive encoding strategy that automatically selects the most suitable time-aware representation for a distinct deployment scenario.
    \item A comprehensive evaluation of \athena on three diverse benchmark IoT datasets, going beyond prior evaluations on a single dataset~\cite{splitech_paper} and demonstrating the system's robustness and generalizability.
    \item An extensive comparative study against traditional positional encodings, feature-based ML algorithms, state-of-the-art time-aware encodings, and recently proposed early detection architectures, establishing the performance advantages of the proposed approach.
    \item A detailed ablation study that quantifies the performance gains contributed by the system's core components and assesses the effect of applying quantization.
\end{itemize}

%% file: sections/2_back_related.tex
Recent advancements in deep learning have significantly improved the capabilities of intrusion detection systems, enabling more accurate and adaptive threat detection. However, challenges remain in achieving efficient and early detection, particularly in resource-constrained environments such as IoT networks. This section reviews deep learning-based IDS, positional encoding mechanisms, and augmentation techniques for cybersecurity, highlighting the gaps that motivate our approach. It also situates the baseline methods evaluated in Section~\ref{sec:results} within their conceptual context, enabling the reader to understand how each relates to \athena's design choices.

\subsection{ML-Based Network Intrusion Detection}
\label{sec:back_related:ids}

Intrusion detection systems leveraging deep learning have gained significant traction in recent years. The predominant approach in the literature relies on extracted features---whether manually engineered statistical properties or automatically learned representations---that abstract away from raw packet content. A smaller but growing subset of recent work processes raw packet bytes directly as model input, eliminating intermediate feature extraction stages entirely.

\subsubsection{Feature-Based Approaches}
\label{sec:back_related:ids:ml}

Feature-based IDS rely on domain expertise to extract statistical properties from network traffic. Traditional approaches employ classical classifiers including Support Vector Machines~\cite{abrar2020icosec, almotairi24ssce}, Random Forests~\cite{le2022sustainability, shafiq21iotj}, Naive Bayes~\cite{almotairi24ssce, shafiq21iotj}, k-Nearest Neighbors~\cite{abrar2020icosec, khan21sensors}, and Decision Trees~\cite{khan21sensors, shafiq21iotj}, often as ensembles~\cite{almotairi24ssce, le2022sustainability}. Recent work has shifted toward deep learning with Multi-Layer Perceptrons~\cite{almohaimeed24applied, farhana20ijece, toupas19icmla}, sequence models like LSTMs and GRUs~\cite{khan21sensors}, autoencoder ensembles~\cite{mirsky18ndss}, and transformer-based architectures~\cite{ferrag2024access}.

While these methods achieve high accuracy, they face critical limitations. Feature extraction introduces computational overhead that delays threat response, depends heavily on domain expertise that may miss novel attack patterns~\cite{khan21sensors, mirsky18ndss}, and most critically for early detection, requires observing substantial session portions to compute meaningful statistics. This creates inherent tension between statistical confidence and detection earliness, as recent IoT-specific work demonstrates: lightweight feature extraction suitable for resource-constrained devices sacrifices the richness needed for early classification, even with sophisticated architectures~\cite{ferrag2024access}. Unlike these feature-engineering-centric approaches, \athena processes raw packet bytes directly, eliminating intermediate extraction stages and the associated latency, enabling classification from the earliest packets of a flow without waiting for sufficient data to compute statistical summaries.

\subsubsection{Raw Traffic Processing}
\label{sec:back_related:ids:raw}

To address the limitations of feature-based representations, recent research has increasingly focused on directly processing raw network traffic, enabling deep models to autonomously learn discriminative representations without relying on domain-specific feature engineering. These approaches employ various deep learning architectures---ranging from CNNs and RNNs to Transformer-based and hybrid models---and can be broadly divided according to their analytical granularity.
Flow-level or session-level models process entire sequences of packets as unified entities, capturing long-range dependencies and temporal correlations within complete communication sessions~\cite{wang21fcnn, pcnn19access, shahhosseini22jnsm, han23compsec, wang18access, zhu21medes}. In contrast, packet-level models operate on individual packets or short, fixed-size windows~\cite{domi2025csr, ogonowski2025cs, zhang19access}, focusing on local byte patterns and short-term statistical cues.

While packet-level models often achieve fine-grained inspection and reduced latency, they may overlook contextual dependencies that span multiple packets, limiting their ability to identify stealthy or evolving threats. Flow-level approaches, by aggregating packets into temporally coherent sequences, capture richer behavioral patterns and enable early yet informed classification decisions---striking a balance between granularity and temporal context. For these reasons, this work adopts a flow-based perspective, emphasizing the analysis of packet sequences to exploit both temporal structure and content dynamics within network flows.

\subsubsection{Models for Early Intrusion Detection}
\label{sec:back_related:ids:early}

EIDS aim to identify malicious behavior early within a network session, enabling faster response and mitigation of potential damage~\cite{lopez19nca}. Unlike conventional IDS that rely on full-session analysis, EIDS seek to classify threats based on incomplete traffic observations, a requirement particularly critical in IoT environments where delayed detection can result in device compromise or large-scale propagation. The fundamental challenge lies in balancing accuracy and earliness: decisions made from partial flows accelerate reaction time but may reduce confidence, while waiting for complete flows increases latency and risk.

Recent advances in deep learning have introduced diverse strategies for achieving early detection, spanning both feature-based methods and raw traffic models. Methods such as convolutional autoencoders~\cite{lunardi2022arcade}, recurrent and attention-based architectures~\cite{djaidja2024early}, adversarial learning frameworks~\cite{kim2022early}, and real-time deep architectures~\cite{callegari2024real} have all been employed. Despite their methodological differences, these studies share a common goal: reducing detection latency while maintaining robust performance under incomplete flow information.

In parallel, a few studies have explored training models on variable-length network flows to enable classification at different stages of a session, highlighting the potential of adaptive detection mechanisms that incrementally refine their predictions as additional packets are observed. Among these, eRNN~\cite{ahmad22icstw} and eTransformer~\cite{ahmad23icstw} apply recurrent and attention-based architectures, respectively, to variable-length raw packet sequences, while eAtt~\cite{islam23cloudcom} and eGlo~\cite{ahmad24icstw} introduce lightweight CNN variants for the same task. Although these architectures demonstrate the feasibility of early classification from raw bytes, none incorporate packet timestamps into the encoding or training process, leaving inter-arrival timing information unexploited. Unlike these prior architectures, \athena jointly addresses temporal representation, early detection optimization, and resource efficiency---providing a unified framework rather than isolated solutions to each sub-problem.

\subsection{Positional Encoding Mechanisms}
\label{sec:back_related:encodings}

Unlike recurrent models that inherently capture sequential order, Transformers rely on fully parallelized self-attention without an intrinsic notion of token order. Positional encodings explicitly encode element positions, enabling the model to learn order-dependent representations while maintaining parallelization advantages. In this work, we consider two encoding categories: (a)~input positional encodings, which are directly added to the input embeddings before processing, and (b)~attention positional encodings, which are applied to the query and key matrices during self-attention computation. Each category serves a distinct purpose in encoding positional information, influencing how the model attends to different parts of the input sequence.

\subsubsection{Traditional Positional Encodings}
\label{sec:back_related:encodings:trad}

Traditional positional encodings were developed primarily for natural language processing tasks, where tokens appear in discrete, evenly spaced positions within a sequence. These methods assign positional information based on integer indices, providing the Transformer with explicit knowledge of element order. The three most widely adopted approaches---sinusoidal, Fourier-based, and rotary encodings---differ in their mathematical formulations and parameter requirements, yet all have proven effective across various sequence modeling tasks. In addition, several alternative non-time-aware mechanisms have been proposed to encode position through learned representations or structural operations.

\paragraph{Sinusoidal Positional Encoding}
The sinusoidal positional encoding was introduced in the original Transformer architecture~\cite{vaswani17nips} as a method to inject position-dependent information into the model without relying on learned embeddings. These encodings are computed using sine and cosine functions of varying frequencies, ensuring that each position has a unique representation while also allowing the model to generalize to unseen sequence lengths. Given a sequence of length $n$, the sinusoidal positional encoding assigns each position $pos \in \{0, 1, \dots, n{-}1\}$ a vector of dimension $d_{\mathrm{m}}$, where $d_{\mathrm{m}}$ denotes the hidden dimension of the Transformer model. The encoding is computed using the following formulas:
\begin{displaymath}
    PE(pos, 2i) = \sin \left( \frac{pos}{10000^{\frac{2i}{d_{\mathrm{m}}}}} \right)
\end{displaymath}
\begin{displaymath}
    PE(pos, 2i{+}1) = \cos \left( \frac{pos}{10000^{\frac{2i}{d_{\mathrm{m}}}}} \right)
\end{displaymath}
where $i = 0, 1, \dots, d_\mathrm{m}/2{-}1 $ indexes the sine and cosine components within the encoding dimension. The choice of the base 10,000 ensures a smooth distribution of frequencies across the encoding dimension, allowing the model to capture both fine-grained and long-range positional dependencies while maintaining numerical stability during training.

\paragraph{Fourier-Based Positional Encoding}
Since the introduction of the sinusoidal encoding, several alternative methods have been proposed to enhance the effectiveness of positional information in Transformers. The Fourier-based positional encoding~\cite{li21nips} extends the idea of the sinusoidal encoding by leveraging a more general Fourier feature mapping. Instead of using a fixed base, this encoding is derived from a learnable frequency basis that enables richer and more flexible positional representations. This approach has been shown to improve the generalization of positional information in various applications. The Fourier positional encoding at position $pos$ is given by:
\begin{displaymath}
    PE(pos, 2i) = \sin(2\pi f_i \, pos)
\end{displaymath}
\begin{displaymath}
    PE(pos, 2i{+}1) = \cos(2\pi f_i \, pos) 
\end{displaymath}
where $f_i$ is the learnable frequency parameter associated with the $i$-th sine-cosine pair in the encoding.

\paragraph{Rotary Positional Encoding}
Another widely adopted method for infusing positional information in Transformer architectures is the rotary positional encoding (RoPE)~\cite{su24neurocomputing}. RoPE introduces position information by applying deterministic rotations directly to the query and key vectors within each attention head. This rotation-based formulation enables the attention mechanism to capture relative positional relationships and supports efficient extrapolation to longer sequences. Let $\mathbf{x} = (x_0, x_1, \dots, x_{d_\mathrm{h}-1})$ denote a query or key vector, where $d_\mathrm{h}$ is the head dimensionality. For position $pos$ and index $i$, RoPE rotates each consecutive pair of components $(x_{2i}, x_{2i+1}) \in \mathbb{R}^2 $ according to:
\begin{displaymath}
    \begin{bmatrix} x_{2i}^{\mathrm{rot}} \\ x_{2i+1}^{\mathrm{rot}} \end{bmatrix} =
    \begin{bmatrix} \cos(pos\,\omega_i) & -\sin(pos\,\omega_i) \\ \sin(pos\,\omega_i) & \cos(pos\,\omega_i) \end{bmatrix}
    \begin{bmatrix} x_{2i} \\ x_{2i+1} \end{bmatrix}
\end{displaymath}
where the rotation frequencies are given by:
\begin{displaymath}
    \omega_i = 10000^{-2i/d_\mathrm{h}}
\end{displaymath}

\paragraph{Additional Non-Time-Aware Encodings}
Beyond the three encodings above, several other non-time-aware mechanisms serve as useful baselines. \emph{Embedding-based} encodings map integer position indices to dense vectors learned during training, providing data-driven positional representations at the cost of additional parameters and potential overfitting in compact models. \emph{Convolutional} encodings employ one-dimensional convolutional filters over the input sequence to extract local positional dependencies dynamically, offering adaptivity but at higher parameter cost. \emph{Global relative} encodings, following the frameworks of Shaw~et~al.~\cite{shaw18acl} and Huang~et~al.~\cite{huang19iclr}, incorporate relative rather than absolute positional information directly into the attention computation by modifying the attention score as:
\begin{displaymath}
    \mathbf{A} = \operatorname{softmax}\left( \frac{\mathbf{Q} \mathbf{K}^\top + \mathbf{Q} \mathbf{E}_\mathrm{r}^\top}{\sqrt{d_{\mathrm{h}}}} \right)
\end{displaymath}
where $\mathbf{E}_\mathrm{r} \in \mathbb{R}^{N \times d_{\mathrm{h}}}$ is a learnable relative position embedding matrix. While all of these non-time-aware methods have proven effective in NLP and other settings, they share a common limitation when applied to network traffic: they treat packet sequences as uniformly spaced, discarding the inter-arrival timing information that often distinguishes malicious from benign behavior. Unlike these approaches, \athena's time-aware encodings replace discrete indices with continuous timestamps, directly capturing temporal dynamics without introducing additional model parameters.

\subsubsection{Time-Aware Positional Encodings}
\label{sec:back_related:encodings:time}

Conventional positional encodings use integer indices $pos \in \{0, 1, \ldots, n{-}1\}$, implicitly assuming uniformly sampled sequences where inter-element intervals are constant or negligible. In contrast, network traffic is inherently a non-uniform time series, with inter-packet arrival times carrying valuable temporal cues. Replacing actual timestamps $t_{pos}$ with discrete indices introduces temporal aliasing, constraining the model's ability to represent timing dynamics. For instance, a brute-force attack with 0.01-second inter-arrival times and benign traffic with 2.0-second intervals receive identical positional vectors under traditional encodings, making them temporally indistinguishable. This information loss prevents the model from leveraging timing irregularities that often characterize malicious behavior.

\paragraph{Applications Across Diverse Domains}
Recent advances across multiple domains have demonstrated the value of time-aware positional encodings for irregular or continuous temporal data. One natural extension adapts sinusoidal encodings to incorporate timestamp differences $\Delta t$, enabling the model to learn periodic yet time-dependent representations~\cite{sharma23icmla, ryu23sera}. Similarly, time-aware rotary encodings extend RoPE with phase shifts proportional to elapsed time, showing improvements on irregular time-series forecasting tasks~\cite{tseriotou2024tempoformer}. These approaches validate the core intuition behind \athena's time-aware sinusoidal and rotary mechanisms. An alternative strategy treats time as a continuous variable and learns its embedding directly. Some methods employ linear mappings---notably CTLPE~\cite{kim2024arxiv}, which adopts a purely linear formulation $PE(pos,:) = a \cdot t_{pos} + b$ ($a, b \in \mathbb{R}^{d_\mathrm{m}}$ learnable), constituting the minimal-complexity continuous-time representation but potentially lacking the capacity to capture cyclical and bursty temporal structure. Others discover mixed sinusoidal representations that adapt to varying periodicities in the data~\cite{bui2025eusipco}. By letting the model determine suitable temporal functions, these approaches offer flexible embeddings for irregularly spaced events without imposing predetermined frequency constraints.

Hybrid and compositional strategies explicitly fuse multiple temporal signals. For example, ChronoFormer~\cite{zhang2025arxiv}, designed for clinical event modeling, employs an additive scheme by summing an absolute sinusoidal timestamp representation with an MLP-based relative inter-arrival mapping: $PE(pos,:) = PE_{\mathrm{abs}}(t_{pos}, :) + PE_{\mathrm{rel}}(\Delta t, :)$, where $\Delta t = t_{pos} - t_{pos-1}$. Packet order and timestamp information can also be merged as $\sin (pos+\log \Delta t)$~\cite{zhou22ecr}. These methods aim to capture both sequential structure and timing patterns, closely mirroring the motivations behind our hybrid encoding framework. This idea has been further refined through learnable weighting schemes: FATA~\cite{zhang2023fata}, designed for sequential tabular data, computes a temporal position as a learnable linear combination $tpos = w_{\mathrm{pos}} \cdot pos + w_{\mathrm{t}} \cdot t_{pos} + b$, where $w_{\mathrm{pos}}$, $w_\mathrm{t}$, and $b$ are trainable parameters, and feeds the result to standard sinusoidal functions, allowing the model to learn the relative importance of sequence order versus actual time intervals. A contrasting line of work avoids mixing these signals at the input stage altogether. PEA~\cite{morenocartagena2023icml}, explored in the context of astronomical light curve analysis, computes content-only attention and adds a time-aware sinusoidal encoding only to the final encoder output, \textit{i.e.}, $\mathrm{output} = z + PE(T)$, cleanly separating content-driven and time-driven reasoning. This design offers competitive performance while reducing training time.

More architecturally advanced approaches embed temporal reasoning directly into the model rather than relying solely on explicit positional encodings. Examples include coupling Transformers with neural ordinary differential equations to model continuous-time evolution~\cite{song2025trajgpt, ji25appint}, fusing Time2Vec embeddings with ODE-based architectures~\cite{fan2025super}, and introducing non-stationary kernels or feature-dependent temporal components~\cite{xiao2025xtsformer, ma22sigkdd}. While these methods achieve strong accuracy on complex temporal modeling tasks, their substantial computational overhead renders them impractical for resource-constrained environments such as IoT gateways and edge devices.

Unlike the methods reviewed above---which either retain discrete indices alongside temporal signals (FATA, ChronoFormer), adopt minimal-complexity linear mappings (CTLPE), or decouple temporal injection from attention entirely (PEA)---\athena directly replaces discrete indices with continuous timestamps, eliminating temporal aliasing, and consistently applies this principle across multiple encoding schemes (sinusoidal, Fourier, and rotary).

\paragraph{Applications in Intrusion Detection}
Within the intrusion detection domain, explicit timestamp modeling remains relatively uncommon. Early work by Han~et~al.~\cite{han23compsec} demonstrated that incorporating $\Delta t$ as explicit features improves detection performance through GTID, which augments sinusoidal encodings with a composite position index $pos' = pos + \log_2(t_{pos}/a+1)$ (where $a$ defaults to $10^{-7}$), merging sequential order and temporal magnitude as complementary features---but, like FATA, retains the discrete index as a signal component. More recently, Miyamoto~et~al.~\cite{miyamoto2024acsac} demonstrated that applying Time2Vec~\cite{kazemi19arxiv} embeddings to network traffic yields measurable improvements over order-only baselines. Time2Vec is a model-agnostic timestamp encoding composed of one linear term capturing global temporal trends and multiple sinusoidal terms modeling periodic patterns:
\begin{displaymath}
PE(:\,,0) = \omega_0 \cdot T + \phi_0, \qquad
PE(:\,,i) = \sin(\omega_i \cdot T + \phi_i),
\end{displaymath}
where $i = 1, \dots, d_{\mathrm{m}}{-}1$ indexes the encoding dimension, offering strong expressive power with minimal built-in inductive bias. However, these studies evaluate isolated encoding formulations without systematic comparison across different time-encoding families.

In contrast to both GTID and Time2Vec, \athena replaces discrete indices with temporal signals and embeds time-aware encoding within an end-to-end framework that incorporates an earliness-oriented training objective and raw-traffic augmentation—treating temporal representation as an integrated component of a unified system rather than an isolated module, while extending our prior work on multi-encoding evaluations for network traffic~\cite{splitech_paper} into a more comprehensive and efficient framework for lightweight intrusion detection in IoT environments.

\subsection{Augmentation Strategies for Cybersecurity Datasets}
\label{sec:back_related:aug}

The availability of real-world cybersecurity datasets is limited, with most existing datasets containing a restricted number of attack sessions, resulting in an insufficient sample size for effective model training. Furthermore, these datasets often include a diverse range of attack types, whose variability and heterogeneity contribute to substantial class imbalances, as certain attack categories may be significantly underrepresented compared to those that generate higher volumes of traffic. To address these challenges, augmentation techniques are essential for enhancing dataset diversity and mitigating class imbalance. By synthetically increasing the number of attack samples, data augmentation improves model generalization, enabling intrusion detection systems to effectively identify both prevalent and rare attack patterns.

To make cybersecurity datasets more varied and balanced, recent studies have explored different augmentation techniques. Nevertheless, the proposed models in these studies predominantly operate on extracted feature representations rather than raw network traffic, leading to the adoption of augmentation techniques primarily designed for tabular data. These approaches include the Synthetic Minority Oversampling Technique (SMOTE)~\cite{mohammad24systems, menssouri2025icc, zhang22fgcs}, Conditional Tabular Generative Adversarial Networks (CTGANs)~\cite{alabdulwahab24sensors, menssouri2025icc}, Variational Autoencoders (VAEs)~\cite{liu22reliability, zhang22fgcs}, Long Short-Term Memory (LSTM) networks~\cite{hasibi2019arxiv}, as well as more advanced methods, such as Transformer-based generative models~\cite{melicias24access} and diffusion models~\cite{jiang24acs, svaroopan24netdiffus}. These techniques have shown considerable success in boosting tabular data variability and quality. However, the literature is missing similar techniques applied to raw network traffic data.

\subsection{Research Gap}
\label{sec:back_related:gap}

The preceding review reveals several interrelated limitations in existing approaches to ML-based intrusion detection that, taken together, motivate \athena's design.

\paragraph{Temporal Representation}
As discussed in Subsection~\ref{sec:back_related:encodings:time}, current time-aware encodings for intrusion detection either evaluate isolated formulations without systematic comparison~\cite{han23compsec, miyamoto2024acsac} or retain discrete indices alongside temporal signals, introducing aliasing artifacts (GTID, FATA). More expressive alternatives, such as Time2Vec and ChronoFormer, have not been evaluated in early detection settings, and purely linear representations (CTLPE) may lack the capacity to capture cyclical and bursty temporal structure. While time-aware encodings have demonstrated value across other domains~\cite{sharma23icmla, ryu23sera, tseriotou2024tempoformer, kim2024arxiv, zhang2025arxiv}, no prior work systematically compares across encoding families specifically for early intrusion detection or proposes a unified principle---replacing discrete indices with continuous timestamps---that generalizes across multiple encoding architectures.

\paragraph{Early Detection}
As discussed in Subsection~\ref{sec:back_related:ids:early}, existing early detection architectures (eRNN, eTransformer, eAtt, eGlo)~\cite{ahmad24icstw, islam23cloudcom, ahmad23icstw, ahmad22icstw} demonstrate the feasibility of variable-length flow classification but lack specialized training objectives that prioritize accurate predictions from minimal packet observations. Moreover, architectures such as those proposed in~\cite{lunardi2022arcade, djaidja2024early, kim2022early} impose computational demands unsuitable for edge deployment, and none incorporates inter-arrival timing into the encoding process.

\paragraph{Data Augmentation}
As reviewed in Subsection~\ref{sec:back_related:aug}, existing augmentation techniques remain tailored to feature-based representations~\cite{mohammad24systems, alabdulwahab24sensors, liu22reliability, melicias24access}, lacking mechanisms to diversify raw traffic while preserving protocol semantics and temporal realism.

\paragraph{Integration}
These challenges---temporal representation, early detection optimization, realistic data augmentation, and resource efficiency---remain largely disjoint in current research. By jointly addressing all four within a lightweight, deployable framework, \athena advances toward practical, real-time protection in resource-constrained IoT environments.

%% file: sections/3_approach.tex
Network traffic can be conceptualized as a collection of flows traversing network elements. The term flow has multiple definitions within the Internet community. According to RFC 7011~\cite{rfc7011}, a traffic flow is a set of packets or frames passing through an observation point over a specified time interval. For instance, in a host-based IDS, the observation point is the potential victim, such as an IoT device vulnerable to attacks. Each packet within a flow shares common attributes, with one of the most widely accepted flow definitions being the 5-tuple representation: source and destination IP addresses, source and destination transport layer ports, and the protocol in use.

Early intrusion detection in the context of a flow-level IDS refers to the capability of accurately classifying a network flow as early as possible, using only a partial sequence of packets within the flow. The goal is to minimize the time and data required for threat identification, enabling rapid response and mitigation before an attack fully unfolds. An effective early detection system balances classification accuracy with earliness, ensuring that malicious activities are identified promptly while minimizing false positives and computational overhead. In order to facilitate early and accurate detection, our approach prioritizes four key objectives:
\begin{itemize}
    \item \textbf{Feature-Free Learning:} Unlike traditional approaches that depend on predefined or extracted features, our system operates directly on raw packet data, minimizing preprocessing overhead and enabling the model to learn representations autonomously.
    \item \textbf{Computational Efficiency:} Our model's primary goal is to remain lightweight, ensuring low latency and minimal memory footprint, making it suitable for deployment on resource-constrained environments such as IoT and edge devices.
    \item \textbf{Rapid Threat Detection:} By classifying network flows using only a small fraction of their packets, our system enables early response and mitigation, preventing attacks from fully unfolding.
    \item \textbf{Adaptability to Variable-Length Flows:} The detection mechanism is designed to handle flows of varying durations without requiring fixed-length representations, ensuring flexibility in real-world traffic analysis.
\end{itemize}

\begin{figure}[t]
    \centering
    \includegraphics[width=0.9\textwidth,trim={8.74cm 1.86cm 8.74cm 1.86cm},clip]{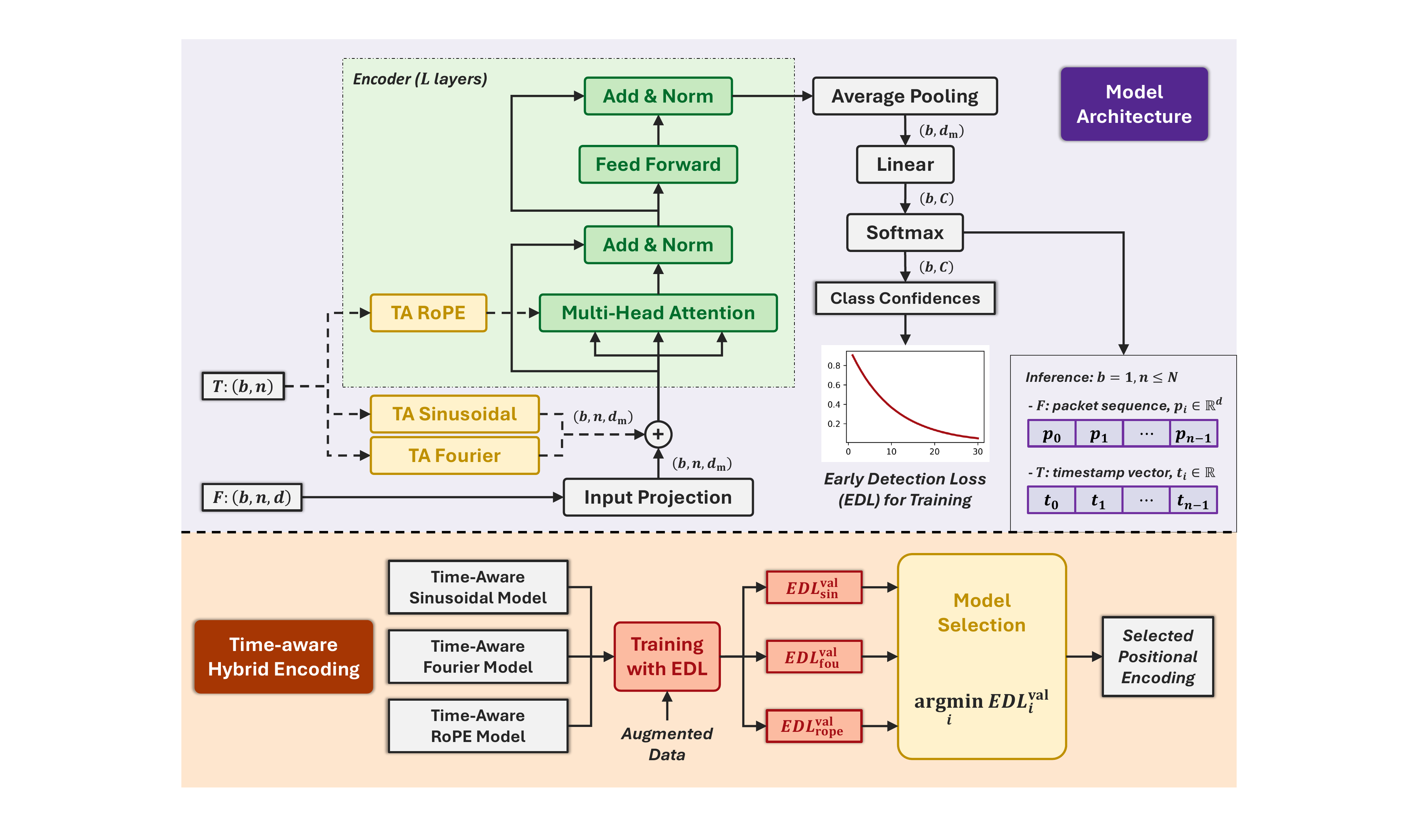}
    \caption{The \athena system architecture. (Top) Transformer encoder ingesting raw packet sequences ($F$) with Time-Aware (TA) encodings from timestamps ($T$): TA Sinusoidal and TA Fourier are added to the input projection, while TA RoPE modifies multi-head attention. The Early Detection Loss (EDL) enforces early classification. (Bottom) The Time-Aware Hybrid Encoding (THE) framework dynamically selects the optimal encoding based on validation loss.}
    \Description{The figure is a two-part block diagram illustrating the Athena system architecture. The top half is titled "Model Architecture." On the left, it takes two inputs: a timestamp vector T with dimensions (b, n), and a packet sequence F with dimensions (b, n, d). The packet sequence F passes through an Input Projection block to become a tensor of dimensions (b, n, d_m). The timestamp vector T is routed to three alternative Time-Aware (TA) positional encodings: TA Sinusoidal, TA Fourier, and TA RoPE. TA Sinusoidal and TA Fourier are shown connecting to an addition node, where they are added directly to the projected packet sequence. In contrast, TA RoPE bypasses this addition and injects directly into the Multi-Head Attention block inside the Encoder. The Encoder block, which has L layers, consists of a Multi-Head Attention module, followed by an Add and Norm module, a Feed Forward module, and a final Add and Norm module. Residual skip connections bypass the attention and feed-forward modules. The output of the encoder feeds sequentially into an Average Pooling layer, yielding dimensions (b, d_m), a Linear layer yielding dimensions (b, C), and a Softmax layer, which outputs "Class Confidences" of dimensions (b, C). For training, these confidences feed into an "Early Detection Loss (EDL)" module, depicted by a graph with a downward-sloping exponential decay curve. An inset box details the inference stage, noting batch size b=1, sequence F as variables p_0 to p_{n-1}, and sequence T as variables t_0 to t_{n-1}. The bottom half is titled "Time-aware Hybrid Encoding." It shows a parallel workflow for three distinct models: a Time-Aware Sinusoidal Model, a Time-Aware Fourier Model, and a Time-Aware RoPE Model. All three undergo "Training with EDL" using "Augmented Data." The training yields three respective validation loss scores: EDL_sin^{val}, EDL_fou^{val}, and EDL_rope^{val}. These three scores feed into a "Model Selection" block which applies an argmin function to find the lowest validation loss. The output of this selection block is the final "Selected Positional Encoding."}
    \label{fig:system}
\end{figure}

Figure~\ref{fig:system} illustrates the core architecture and methodological framework of the proposed A-THENA system. To achieve the aforementioned objectives, the model directly ingests raw packet flows ($F$) alongside their corresponding continuous arrival timestamps ($T$). As depicted in the top panel, raw byte sequences are projected into a hidden feature space, where temporal dynamics are injected to establish sequential order and inter-arrival timing. Depending on the specific variant, the timestamps are transformed via Time-Aware (TA) Sinusoidal or Fourier encodings and added directly to the input embeddings, or applied as a TA Rotary Positional Encoding (RoPE) directly within the Transformer encoder's Multi-Head Attention mechanism. The encoded sequence is then aggregated via average pooling and passed through a classification head to produce class confidences. Crucially, the network is optimized using the Early Detection Loss (EDL), which applies an exponentially decaying penalty to enforce accurate classification from the earliest possible packets. Finally, the bottom panel outlines the Time-Aware Hybrid Encoding (THE) framework: to adapt to diverse network environments, the system evaluates all three time-aware models and dynamically selects the optimal configuration by minimizing the validation loss ($\text{argmin}_i EDL_i^{\text{val}}$).

\begin{figure}
    \centering
    \includegraphics[width=\textwidth,trim={6.38cm 1.86cm 6.38cm 1.84cm},clip]{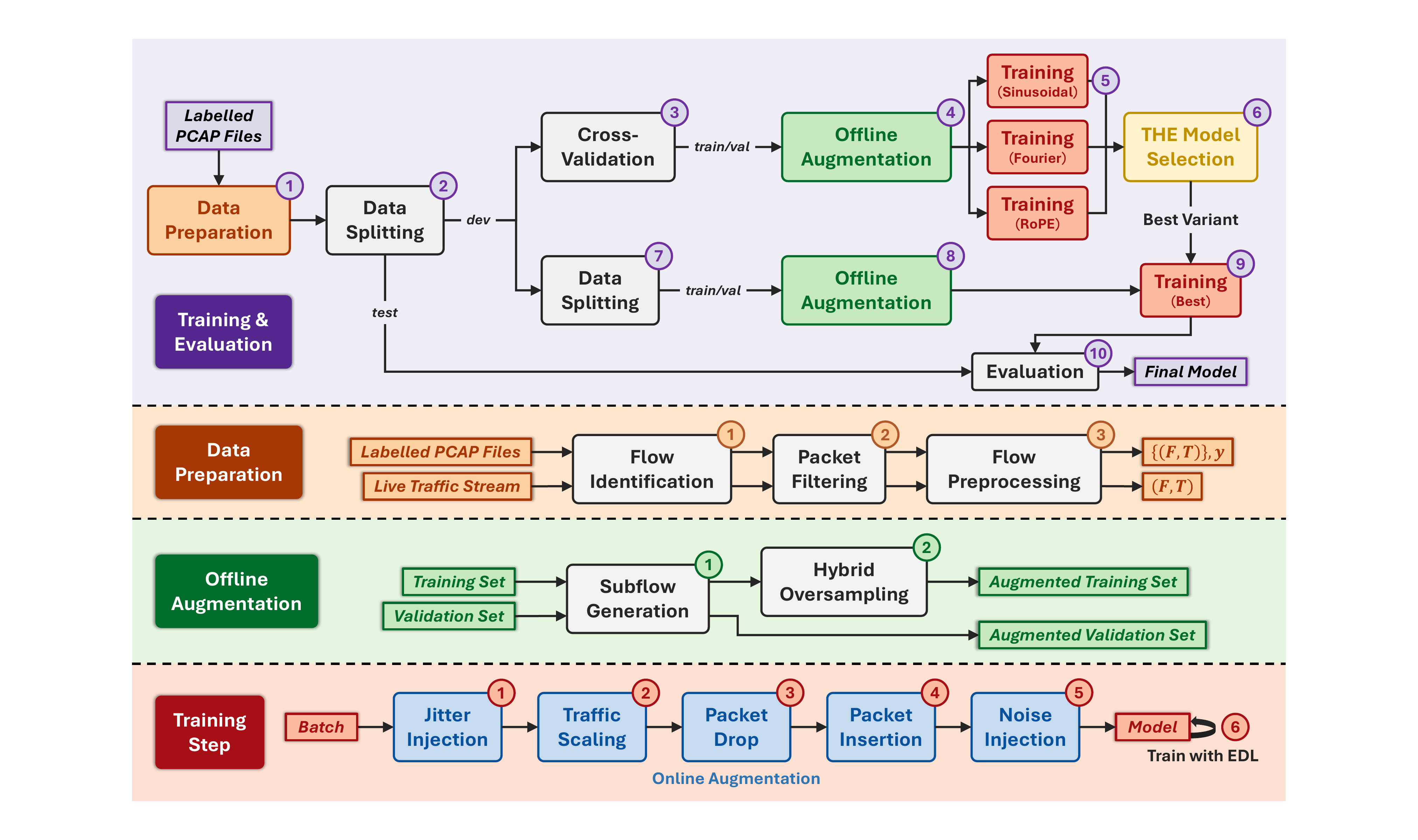}
    \caption{\athena's \textit{Training \& Evaluation} workflow. Modules are color-coded by functional role: data preparation (brown), model creation (purple), offline augmentation (green), THE variant selection (yellow), online network-specific augmentation (blue), and the EDL-based optimization step (red). Numbered circles denote the execution order within each stage, and arrows show data flow.}
    \Description{This diagram illustrates the overview of the proposed system's offline workflow for building the early intrusion detection model. Labeled PCAP files are processed through the Data Preparation pipeline—comprising Flow Identification, Packet Filtering, and Packet Preprocessing—to produce tuples {(F, T)}, where F denotes normalized packet bytes and T denotes timestamp vectors; the same pipeline can be applied to live traffic to generate (F,T) tuples for inference. The dataset is then split into test and development partitions, with the latter undergoing cross-validation. Each fold's training and validation subsets are augmented via Offline Augmentation, which includes Subflow Generation for early-stage flow simulation and Hybrid Oversampling for class balancing. Three models are subsequently trained using different temporal encoding variants (Sinusoidal, Fourier, RoPE) within the THE framework, optimized via EDL and supported by Online Augmentation (Jitter Injection, Traffic Scaling, Packet Drop, Packet Insertion, Noise Injection). The best-performing variant is selected and retrained on the full development set before final evaluation on the test set. The resulting model can be deployed for real-time inference, continuously monitoring live network traffic, with offline evaluation indicating expected deployment performance.}
    \label{fig:workflow}
\end{figure}

To operationalize this architecture, Figure~\ref{fig:workflow} details the end-to-end data processing and experimental workflow used to construct and evaluate the model. Labeled PCAP files are first processed through the \textit{Data Preparation} pipeline to produce structured flow representations and corresponding labels; this same pipeline is applied to live traffic to generate equivalent representations during real-time inference. After data preparation, the dataset is split into a hold-out test set and a development set undergoing cross-validation. Training and validation subsets are processed by the \textit{Offline Augmentation} module for early-stage simulation and class balancing. During each training fold, batches are further perturbed through \textit{Online Augmentation}, and the three temporal encoding variants are trained in parallel. Once the THE framework identifies the best-performing encoding based on validation performance, a final training run is conducted on the full development set. The resulting model is then ready for deployment and offline evaluation. The subsequent subsections detail the data preparation pipeline (Section~\ref{sec:system:data_preparation}), the specialized augmentation and training strategies (Section~\ref{sec:system:training}), and the formal definition of the system architecture and encodings (Section~\ref{sec:system:model}).

\subsection{Data Preparation}
\label{sec:system:data_preparation}

Our approach maximizes computational efficiency by directly processing raw packet bytes instead of relying on predefined features. This eliminates preprocessing overhead, reduces latency, and removes the need for manual feature engineering, allowing the model to learn meaningful representations autonomously. Additionally, it preserves fine-grained information that traditional feature extraction might discard. The following subsections outline the three key modules of this process.

\subsubsection{Flow Identification}

The initial step in the proposed system involves maintaining a record of all active flows within the network, regardless of whether they originate from network captures (PCAP files) or real-time network traffic. Upon the arrival or departure of a new packet, the system determines whether the packet corresponds to an existing flow or necessitates the creation of a new flow instance.

Formally, a flow $F$ can be represented as an ordered sequence of packets:
\begin{equation}
    F = ( p_0, p_1, \dots, p_{n-1} ),
    \label{eq:flow_def}
\end{equation}
where $p_i \in \mathbb{R}^d$ is the $i$-th packet, $d$ is the length of a packet in number of bytes, and $n$ is the length of the flow. In our system, each flow can have a maximum length $N$, therefore, the number of packets in any given flow satisfies the condition $1 \leq n \leq N$.

To construct network flows from raw traffic, a well-defined methodology must be established. Network attacks exhibit diverse characteristics, behaviors, and patterns. For the conventional 5-tuple definition---which consists of the source IP, destination IP, source port, destination port, and protocol---to be effective in attack detection, these parameters must remain consistent throughout the entire attack session. For instance, brute force attacks (\textit{e.g.}, SSH, FTP, or RDP brute forcing) involve repeated login attempts to a specific service, maintaining a stable 5-tuple across multiple authentication requests. However, many sophisticated cyber threats do not adhere to a consistent 5-tuple structure, rendering detection based solely on this definition inadequate. Examples include distributed attacks, where IP addresses frequently change, and scanning attacks or port-hopping techniques, which result in a high number of ephemeral, often two-packet flows, if classified using the 5-tuple. These limitations prevent an effective representation of the ongoing attack.

To address this challenge, the system begins with Level 0 aggregation, corresponding to the standard 5-tuple definition, which is sufficient for most single-session attacks. The system also tracks the number of active flows and periodically compares it against a threshold defined as five times the normal traffic baseline (\textit{i.e.}, the average number of active flows observed during benign operation). If this threshold is exceeded, the system escalates to Level 1 aggregation, where flows are grouped using the 3-tuple (source IP, destination IP, protocol), effectively ignoring transport-layer ports. If the number of aggregated flows remains above the threshold, the system advances to Level 2 aggregation, defined by the 2-tuple (destination IP, protocol); at this stage, the destination IP corresponds to the victim host receiving the attack traffic. Once traffic levels fall back below the threshold, the system gradually returns to Level 0. This adaptive strategy maintains lightweight flow tracking during normal conditions while expanding aggregation only when necessary, thereby improving scalability and preserving detection effectiveness under high-load or attack scenarios.

\subsubsection{Packet Filtering}

The second stage in data preparation involves isolating network traffic relevant to the specific threat models and topology targeted by the IDS. By discarding irrelevant data, this module significantly reduces computational overhead and focuses the model's attention on high-risk interactions. The filtering process operates on two distinct levels: protocol selection and network-specific constraints.
\begin{enumerate}
    \item First, the system filters based on protocol susceptibility. Services such as HTTP, ARP, and ICMP are frequently exploited, making them critical candidates for analysis. For instance, retaining HTTP traffic allows the system to inspect client-server exchanges for web-based threats, including Cross-Site Scripting (XSS), SQL injection, and Denial-of-Service (DoS) attacks.
    \item Second, the module accommodates network-specific context to target precise communication channels. This allows for the inclusion of filtering rules based on known network topology or specific attack scenarios. For example, in the context of a brute-force attack targeted at a specific server, the filter can be configured to isolate traffic exclusively between the attacker's suspected subnet and the victim's IP address. By narrowing the scope to these specific source-destination pairs, the system effectively eliminates background noise, ensuring that the deep learning model processes only the traffic pertinent to the anticipated attack vector.
\end{enumerate}

\subsubsection{Flow Preprocessing}

Flow preprocessing constitutes the final stage of the data pipeline, essential for transforming raw network traffic into structured arrays compatible with deep learning models. This phase encompasses data cleaning, byte-level normalization, and the standardization of inputs into fixed-sized arrays.

Processing begins at the packet level by stripping superfluous metadata. The Ethernet header is discarded entirely, while the source and destination addresses are removed from the IP header. Eliminating these specific addresses prevents the model from overfitting to environmental artifacts, thereby encouraging it to focus on protocol behavior and payload content rather than network topology. Subsequently, each packet is either truncated or padded to a uniform length, denoted as $d$ in Equation~\eqref{eq:flow_def}. To optimize gradient descent and ensure training stability, byte values are normalized to the range $[0,1]$ by dividing by $255$. This scaling step effectively mitigates numerical disparities in the input data, facilitating faster convergence~\cite{huang2020pami}.

To capture the temporal dynamics of the flow, packet timestamps are extracted and stored as a separate vector,
\begin{equation}
    T = (t_0, t_1, \dots, t_{n-1}),
    \label{eq:timestamps}
\end{equation}
which serves as a secondary input to the system. The timestamp of the first packet in a flow is set to $0$, and subsequent timestamps represent the absolute time elapsed since the first packet's arrival. This temporal information is vital for recognizing sequential patterns in network traffic and detecting anomalies based on timing deviations.

Finally, to accommodate the fixed-length input requirements of Transformer architectures, all flows must be standardized to a uniform sequence length $N$. Flows containing fewer than $N$ packets are padded with zero-valued vectors, and their corresponding timestamp sequences are similarly zero-padded. In conjunction with this process, binary attention masks are generated to differentiate between authentic data and padding. Valid positions are assigned a mask value of $1$, while padded regions are assigned $0$, ensuring that the model's attention mechanism focuses exclusively on meaningful network activity.

\subsection{Training}
\label{sec:system:training}

The training phase of the proposed system aims to develop deep learning models capable of early attack detection using raw network packet data. This phase follows the data preparation stage, leveraging the preprocessed packet representations and their corresponding timestamp vectors as input.

\subsubsection{Data Splitting \& Cross-Validation}
\label{sec:system:training:data_split}

To ensure a rigorous evaluation, we first isolate a hold-out test set, which remains strictly unseen during the entire training process. The remaining dataset constitutes the development set, which is subjected to $k$-fold cross-validation. Within each fold, the data is dynamically split into training and validation subsets. This approach allows us to tune hyperparameters and assess model stability using the validation splits, while reserving the hold-out test set for the final confirmation of the model's effectiveness on real-world, unseen traffic. After cross-validation identifies the best-performing configuration, the development set is employed to train the final model.

\subsubsection{Offline Augmentation}
\label{sec:system:training:off_aug}

Building on the offline augmentation stage introduced in our prior work~\cite{splitech_paper}, we substantially enhance its functionality by redesigning subflow generation to be dynamic and class-aware for improved early detection training, and by introducing a hybrid oversampling mechanism that compensates for class imbalance using the resulting subflow distributions.
\begin{enumerate}
    \item \textbf{Subflow Generation:} A subflow is defined as the first $k$ packets of an original flow of length $n$, where $1 \le k < n$. Rather than exhaustively generating all possible subflows---which can create massive data redundancy---we dynamically determine the number of subflows to generate based on the model's complexity and per-class statistics. We define a target sample count per class, $m_\mathrm{d}$, proportional to the model's parameter count ($P$) and the number of classes ($C$), aiming for a density of $m_\mathrm{d} \approx (2P) / C$. For each class $c$ with size $m_c$ and average flow length $n_c$, subflow generation proceeds conditionally:
    \begin{itemize}
        \item \textit{Minority Classes ($m_c < m_\mathrm{d}$):} To compensate for the deficit $m_\mathrm{d} - m_c$, we compute an augmentation factor $a_c = \min\!\big((m_{\mathrm{d}} - m_c) / m_c,\, n_c-1 \big)$. As this factor is typically non-integer, we employ stochastic rounding to determine the exact count per sample: each original flow generates $\lfloor a_c \rfloor$ subflows, with one additional subflow generated with probability $a_c - \lfloor a_c \rfloor$. Cut-off positions $k$ are sampled logarithmically, yielding dense coverage of early packets and sparser sampling toward the tail, an essential property for early-stage detection.
        \item \textit{Majority Classes ($m_c \ge m_\mathrm{d}$):} Since these classes already possess sufficient statistical representation, further bulk augmentation would exacerbate class bias. However, to ensure the model learns to detect these classes early, we define a fixed augmentation factor $a_c = 0.2$; we randomly select $20\%$ of the flows in each class and generate exactly one short subflow (restricted to $k \in [1, 5]$) for each. This prioritizes the injection of early-stage patterns without bloating the class size.
    \end{itemize}
    The resulting per-class sample count after subflow generation is $m'_c \approx (1 + a_c)\, m_c$.
    \item \textbf{Hybrid Oversampling:} While subflow generation increases dataset size, residual imbalance may persist. To guarantee perfect class parity, we apply a hybrid oversampling strategy that combines deterministic coverage with stochastic filling. We define a reference size $m_\mathrm{max} = \max_c (m'_c)$ and calculate the required oversampling factor $z_c = (m_\mathrm{max} - m'_c) / m'_c$ for every class $c$. We decompose $z_c$ into an integer component $r = \lfloor z_c \rfloor$ and a decimal component $p = z_c - r$.
    The oversampling proceeds in two steps:
    \begin{itemize}
        \item \textit{Deterministic Step:} Every sample in class $c$ is duplicated exactly $r$ times. This ensures that the base distribution of the minority class is preserved and uniformly upweighted.
        \item \textit{Stochastic Step:} To account for the remaining fractional deficit, we randomly sample $p \cdot m'_c$ instances from the original set (without replacement within this subset) to receive one additional copy.
    \end{itemize}
    The final sample size for all classes is therefore $m''_c \approx m_{\mathrm{max}}$.
\end{enumerate}

As shown in Figure~\ref{fig:workflow} (highlighted in green), this full methodology targets the training set. The validation set undergoes subflow generation using the same class-specific factors ($a_c$) derived from training statistics. We strictly exclude oversampling to preserve original class proportions, ensuring unbiased evaluation on a representative distribution of partial flows.

\subsubsection{Online Augmentation}
\label{sec:system:training:on_aug}

Our online augmentation stage adopts the packet-level perturbation techniques introduced in our prior work~\cite{splitech_paper} and applies them without modification. As shown in Figure~\ref{fig:workflow} (highlighted in red), each flow in a batch is subjected to stochastic perturbations during training, enhancing robustness and complementing offline oversampling. While oversampling replicates minority flows, the online stage ensures that each replica undergoes a distinct random transformation, preventing the model from seeing identical packets twice and reducing memorization risks. By leveraging packet timestamps, the pipeline also introduces time-aware perturbations that increase variability beyond what time-agnostic systems can achieve.
\begin{enumerate}
    \item \textbf{Jitter Injection:} The first augmentation technique focuses on the timestamps of the packets within a flow. Jitter injection introduces small, random variations in packet arrival times to simulate the real-world jitter commonly observed in network communications. For each timestamp, the minimum temporal distance to adjacent packets, denoted as $ t_{\mathrm{min}} $, is first calculated, and then a random perturbation is sampled from the continuous uniform distribution and applied:
    \begin{displaymath}
        \mathcal{U}(-0.7 \cdot t_{\mathrm{min}}, 0.7 \cdot t_{\mathrm{min}}).
    \end{displaymath}
    This perturbation helps simulate network conditions such as congestion or packet delays, making the model more robust to variations in packet timing.
    \item \textbf{Traffic Scaling:} The second technique applied to the timestamps is traffic scaling. This method simulates different network speeds by randomly choosing a scaling factor from the set $ \{0.5, 0.75, 1.0, 1.25, 1.5\} $. This scaling factor is then applied to the timestamps, either increasing them to simulate slower networks or reducing them to mimic high-speed links. This variation exposes the model to different network conditions and improves its ability to generalize across a wide range of traffic speeds.
    \item \textbf{Packet Drop:} This is the first augmentation technique to operate at the packet level. Packet drop randomly drops a number of packets from each flow. The maximum number of packets that can be dropped depends on the length of the flow, calculated as
    \begin{displaymath}
        \textit{max\_packets\_to\_drop} = \lfloor 0.25 \cdot n - 0.5 \rfloor,
    \end{displaymath}
    where $n$ is the length of the flow. The actual number of packets to drop is drawn from the discrete uniform distribution:
    \begin{displaymath}
        \mathcal{U} \{0, \textit{max\_packets\_to\_drop} \}.
    \end{displaymath}
    \item \textbf{Packet Insertion:} The packet insertion technique is the second packet-level method. It randomly adds a number of zero-byte packets into a flow. The maximum number of zero packets to be inserted is based on the flow length and is calculated as
    \begin{displaymath}
        \textit{max\_packets\_to\_insert} = \lfloor 0.15 \cdot n - 0.5 \rfloor,
    \end{displaymath}
    The actual number is again drawn from a discrete uniform distribution:
    \begin{displaymath}
        \mathcal{U} \{0, \textit{max\_packets\_to\_insert} \}.
    \end{displaymath}
    \item \textbf{Noise Injection:} The final augmentation technique involves adding noise to the bytes of the packets. For each flow, at most $ \lfloor n/3 \rfloor $ packets are modified, and for each modified packet, at most $ \lfloor d/100 \rfloor $ bytes are altered. The positions of the bytes to be modified are randomly selected from a discrete uniform distribution. The noise itself is drawn from a continuous normal distribution with zero mean and a standard deviation of $0.1$. Since all values have already been normalized to the range $[0,1]$, the added noise is small but effective in simulating random variations or errors in the packet data.
\end{enumerate}

It is important to emphasize that not every augmentation technique is applied to every sample. For instance, a scaling factor of $1$ results in no modification, and if no packets are selected for dropping, insertion, or noise injection, those operations remain inactive. Beyond adopting this component as-is, our contribution includes a rigorous evaluation of both the time-aware augmentation techniques and the full online augmentation pipeline through an extensive ablation study across multiple network environments (Section~\ref{sec:results:ablation}).

\subsubsection{Early Detection Loss Function}
\label{sec:system:training:loss_fn}

Aiming to improve early classification performance, as initially proposed in our prior work~\cite{splitech_paper}, we employ the Early Detection Loss (EDL) to emphasize accurate predictions at earlier stages. In this work, we present an extended evaluation of its effectiveness across multiple datasets. EDL applies greater penalties to misclassifications in shorter flows compared to longer ones, thereby encouraging the model to make accurate decisions even with limited packet information---an essential capability for timely anomaly detection in real-time network traffic analysis.

During training, for each batch of $b$ samples, the cross-entropy loss is first computed individually for each sample. The overall batch loss is then obtained as a weighted average of these individual losses:
\begin{equation*}
    L = \sum_{i=0}^{b-1} w_i \, CE_i,
    \label{eq:loss_fn}
\end{equation*}
where $w_i$ represents the weight associated with the $i$-th sample, and $CE_i$ is the corresponding cross-entropy loss. The weight $w_i$ is defined as
\begin{equation*}
    w_i = e^{-0.1 \cdot n_i},
\end{equation*}
where $n_i$ denotes the length of the $i$-th flow in the batch. This weighting mechanism ensures that the model prioritizes minimizing errors in shorter flows, thereby improving its effectiveness in early classification scenarios. By exponentially reducing the weight as $ n_i $ increases, the model is encouraged to make more accurate predictions with fewer packets, which is essential for applications requiring quick, real-time decisions.

\subsection{System}
\label{sec:system:model}

In this work, we adopt the same lightweight Transformer-based architecture as in our earlier work~\cite{splitech_paper} (as shown in Fig.~\ref{fig:system}, top), optimized for low-latency detection on constrained devices. Unlike traditional models, Transformers can effectively learn packet relationships independent of their positions, making them well-suited for attack detection and traffic classification. Additionally, the temporal aspect of packet arrival times can be crucial for identifying malicious activity, as many attacks exhibit distinct timing patterns. The Transformer architecture offers a significant advantage in this regard, as it can leverage the temporal information through time-aware positional encodings, allowing it to model temporal dependencies within a flow. This capability improves the accuracy and robustness of attack detection by leveraging both sequential and temporal information.

\subsubsection{Base Model}
\label{base_model}

Our base model, defined as the core Transformer architecture excluding positional encoding, is intentionally designed for low-latency inference with minimal parameter count, while preserving sufficient representational capacity for classification. Below, we summarize the architecture and hyperparameters for completeness.

Building upon the standard Transformer framework introduced in~\cite{vaswani17nips}, we incorporate specific modifications tailored for the analysis of network flows. As our task is classification, only the encoder component of the Transformer model is employed. Since the preprocessed raw bytes from each packet directly serve as token embeddings, an explicit input embedding layer is not required. However, to align the input data with the model's feature space, we apply a fully connected layer that maps the input dimension $d$ to the hidden dimension $d_\mathrm{m}$. The transformed input is subsequently processed through a sequence of $L$ Transformer encoder blocks, in which we replace the standard GELU activation function with ReLU to enhance computational efficiency and promote training stability. Following the final Transformer encoder block, global average pooling aggregates the sequence of packet-level embeddings into a single fixed-length vector. This aggregated representation is then passed through a fully connected layer comprising $C$ output neurons and is subsequently processed by a softmax activation function to produce class confidence scores.

\begin{table}[t]
    \small
    \caption{System Hyperparameters and their Corresponding Values}
    \label{tab:hyperparams_values}
    \begin{tabular}{rcr}
        \toprule
        \textbf{Hyperparameter} & \textbf{Symbol} & \textbf{Value} \\
        \midrule
        Packet length & \(d\) & 448 \\
        Maximum flow length & \(N\) & 30 \\
        \midrule
        Hidden dimension & \(d_\mathrm{m}\) & 8 \\
        Number of Transformer blocks & \(L\) & 1 \\
        Number of attention heads & \(h\) & 4 \\
        Attention head dimension & \(d_\mathrm{h}\) & 8 \\
        FFN intermediate dimension & \(d_\mathrm{ff}\) & 16 \\
        Dropout rate & \(p_\mathrm{drop}\) & 0.1 \\
        Number of output classes & \(C\) & \textit{dataset-specific} \\
        \bottomrule
    \end{tabular}
\end{table}

Table~\ref{tab:hyperparams_values} summarizes the hyperparameters employed in our system. These values were selected to achieve rapid inference and a lightweight model architecture, essential for efficient operation in practical network environments. Notably, we do not impose the constraint $d_\mathrm{h} = d_\mathrm{m}/h$, providing additional flexibility in configuring attention heads. The choice of values for input data representation was guided by empirical analysis and practical considerations:
\begin{itemize}
    \item A maximum sequence length of $N = 30$ packets allows the model to capture sufficient contextual information while maintaining low computational complexity, considering that most malicious activities manifest within shorter packet sequences.
    \item A packet feature size of $d = 448$ provides comprehensive representation, capturing essential packet information such as critical headers and payload content, necessary for accurate attack detection.
\end{itemize}

The total number of trainable parameters in our base model is calculated as
\begin{displaymath}
    P = d_\mathrm{m} [ 2L ( 2hd_\mathrm{h} + d_\mathrm{ff} + 3 ) + d + C + 1 ] + L( 3hd_\mathrm{h} + d_\mathrm{ff} ) + C,
    \label{eq:param_count}
\end{displaymath}
which indicates that increasing either the model width ($d_\mathrm{m}$) or depth ($L$) significantly increases the model's parameter count, with direct implications for computational overhead and memory usage. With our selected hyperparameters, the base model consists of approximately 5,070 total trainable parameters, excluding any positional encoding mechanisms. This lightweight architecture facilitates rapid decision-making while maintaining the necessary representational power to distinguish between normal and attack traffic.

\subsubsection{Time-Aware Positional Encodings}
\label{sec:system:system:ta_encodings}

As discussed in Subsection~\ref{sec:back_related:encodings}, conventional positional encodings rely on uniformly spaced sequence indices. In network traffic, however, packet flows exhibit non-uniform inter-arrival times, rendering this assumption invalid and limiting the suitability of standard positional encoding schemes. To address this limitation, \athena's encoding mechanism replaces the discrete index vector $(0, 1, \dots, n{-}1)$ with the timestamp vector $T = (t_0, t_1, \dots, t_{n-1})$. This modification is applied to standard sinusoidal, Fourier-based, and rotary encodings to produce three time-aware variants: TA Sinusoidal, TA Fourier, and TA RoPE. These variants were first introduced in our prior work~\cite{splitech_paper} as individual alternatives to traditional encodings, specifically engineered to capture the temporal irregularities inherent in sequential network traffic.

\begin{figure}
    \centering
    \includegraphics[width=0.331\textwidth]{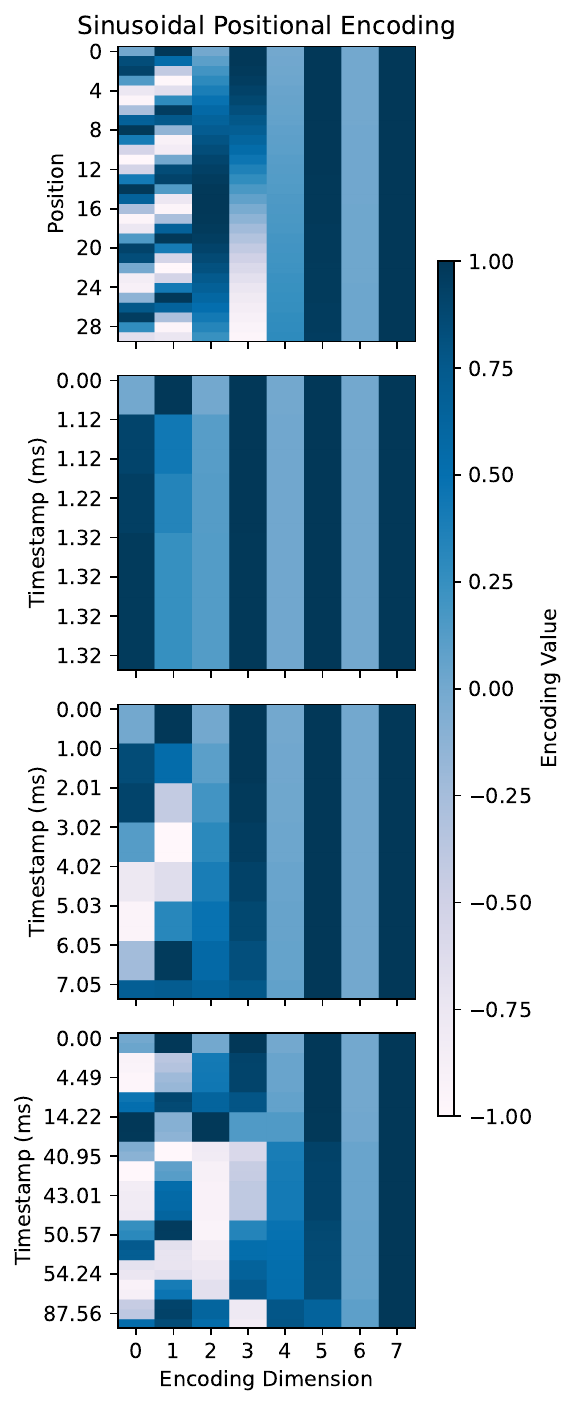}
    \includegraphics[width=0.331\textwidth]{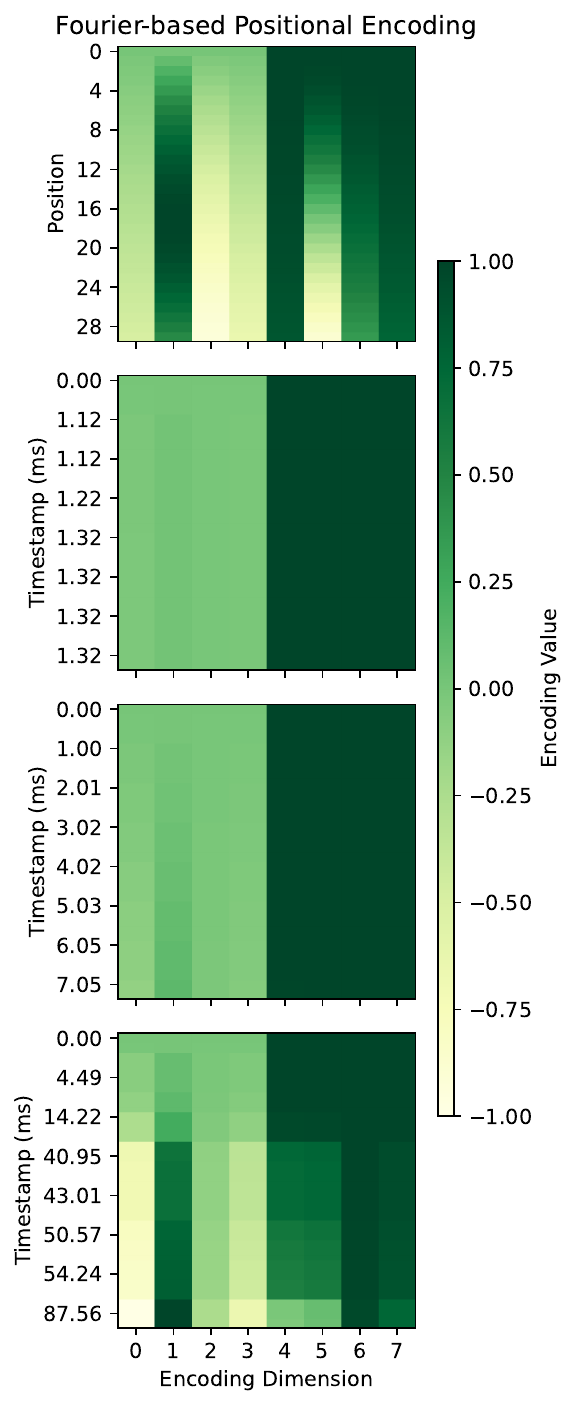}
    \includegraphics[width=0.312\textwidth]{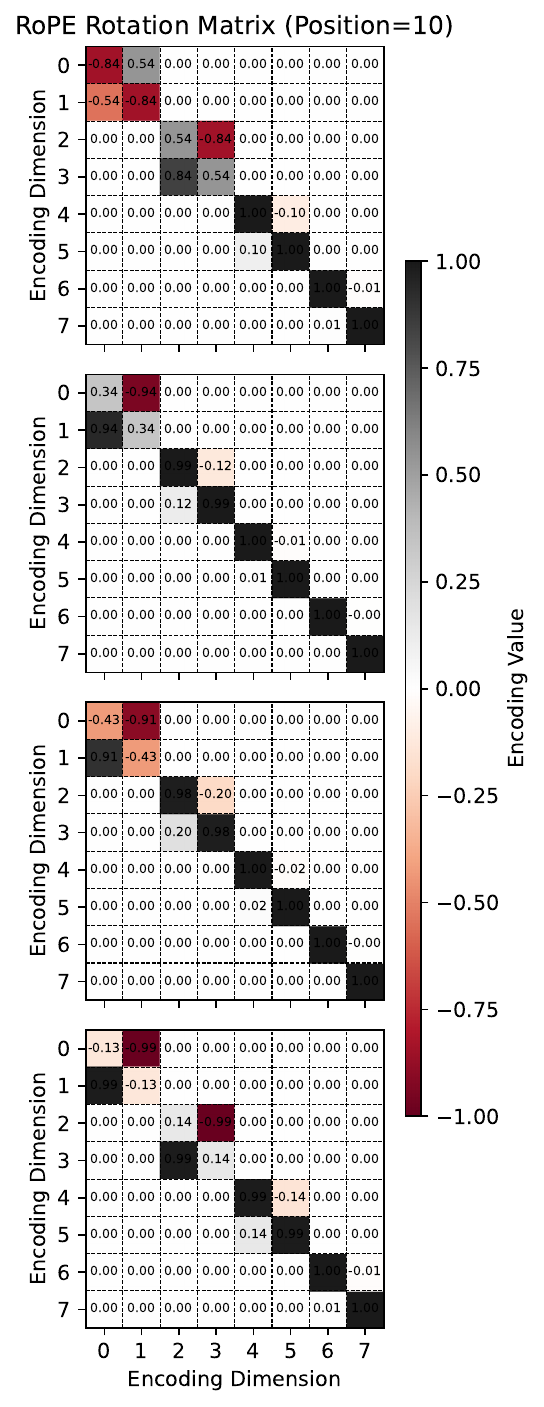}
    \caption{Comparison of standard index-based positional encodings (top row) against the proposed time-aware variants applied to SSH brute-force, benign, and backdoor malware traffic scenarios.}
    \Description{This figure visualizes and compares three types of time-aware positional encodings used in Transformer models: sinusoidal encoding, Fourier-based encoding, and rotary position embeddings (RoPE). Each column corresponds to one encoding type. The left column shows sinusoidal positional encodings as 2D heatmaps across encoding dimensions and input positions or timestamps, with four distinct timestamp ranges plotted vertically. The middle column presents Fourier-based encodings with smoother and more continuous patterns, also across four timestamp intervals. The right column shows RoPE rotation matrices for position 10, visualized as 8×8 matrices of rotation values across encoding dimensions, with numbers overlaid to indicate exact values. Encoding values are color-mapped, ranging from -1 to 1. The horizontal axis for the heatmaps represents encoding dimensions, while the vertical axis shows position or timestamp in milliseconds. The visualizations illustrate how encoding structures vary with time, highlighting differences in temporal sensitivity and expressiveness across the three approaches.}
    \label{fig:ta_encodings}
\end{figure}

Figure~\ref{fig:ta_encodings} demonstrates the capability of the time-aware positional encodings to effectively represent temporal patterns within network traffic, compared to conventional positional encodings. The first row depicts traditional positional encodings, which rely exclusively on discrete packet indices and thus omit temporal context. The subsequent rows introduce the proposed time-aware positional encodings, computed using the timestamps of packets within individual network flows. Specifically, the second row corresponds to a representative SSH brute-force attack scenario, characterized by packets transmitted in rapid succession, thus yielding subtle encoding variations. The third row illustrates benign network traffic with moderately spaced packet transmissions, resulting in clearer encoding distinctions. Finally, the fourth row corresponds to a web-based backdoor malware attack scenario with significantly delayed packet transmissions, exhibiting the most pronounced encoding variations. Across all positional encoding methods considered, the proposed time-aware variants distinctly and systematically reflect absolute temporal differences, thereby highlighting their suitability for accurately capturing critical timing information essential for enhanced network flow classification.

Beyond network traffic analysis, time-aware encodings can be extended to encompass other types of time series data, where samples are generated at irregular intervals, leading to unevenly spaced timestamps. Examples of such data include sensor readings recorded only during significant changes, user activity logs (\textit{e.g.}, clicks, searches, and logins), as well as social media and communication logs, among others. By integrating time-aware encodings, these domains can benefit from more precise temporal representations, improving downstream predictive tasks.

\paragraph{Theoretical Justification}
Traditional positional encodings rely on uniformly spaced indices that fail to capture the non-uniform, irregular nature of packet inter-arrival times found in real-world network traffic. In contrast, time-aware encodings replace fixed indices with actual packet timestamps, providing an input that reflects the temporal structure of a flow. Theoretically, this enhances the Transformer's inductive bias by embedding time directly into the self-attention computation. The dot-product attention mechanism can now weigh packet-packet interactions based not only on content but also on temporal proximity---making it more sensitive to traffic bursts, delays, and timing anomalies. This is especially valuable in intrusion detection, where attacks often exhibit characteristic timing patterns (\textit{e.g.}, periodic scanning, flood attacks). By interpreting timestamp-based encodings as a mapping into a temporal frequency space (via sinusoidal or Fourier functions), we enable the model to capture these temporal dynamics in a principled way, ultimately improving its ability to detect malicious flows early and accurately.

Consider a generic positional encoding function $PE(pos)$ that maps each position to a vector representation.
\begin{itemize}
    \item Under traditional encodings, $pos$ is a discrete integer index. Consequently, the positional difference between any two consecutive packets is always identical ($pos{+}1$ vs.\ $pos$).
    \item Under our time-aware formulation, $pos$ is replaced by the continuous timestamp $t_{pos}$. As a result, the geometric distance between the encodings of two packets becomes a direct function of their actual temporal separation ($\Delta t = t_{pos+1} - t_{pos}$).
\end{itemize}

For example, a brute-force attack might transmit packets with a $\Delta t$ of 0.01 seconds, whereas benign traffic might have a $\Delta t$ of 2.0 seconds. In our model, the encoding vectors $PE(t_{pos})$ and $PE(t_{pos} + 0.01)$ will be very close in the embedding space, whereas $PE(t_{pos})$ and $PE(t_{pos} + 2.0)$ will be significantly more distant. This enables the encoding to act as a frequency-based transformation of the time series of packet arrivals, embedding both fine- and coarse-grained temporal variations. The attention mechanism, in turn, becomes sensitive to timing patterns in network flows, making it theoretically more suited for detecting irregular, bursty, or periodic attack behaviors compared to index-based encodings.

\subsubsection{\athena}
\label{sec:system:system:the}

The proposed system identifies the optimal time-aware positional encoding by evaluating each encoding's effectiveness with respect to the specific characteristics of the attack types analyzed. Among the sinusoidal, Fourier-based, and rotary formulations, the model with the lowest average validation loss across cross-validation folds is selected, ensuring strong generalization to unseen data. This selection process underpins our \textbf{T}ime-Aware \textbf{H}ybrid \textbf{E}ncoding (\textbf{THE}) framework (Fig.~\ref{fig:system}, bottom), which captures the temporal dynamics of network flows through an adaptive combination of multiple encoding strategies. By dynamically choosing the encoding variant best aligned with the observed traffic patterns, THE introduces an adaptive capability not present in prior time-aware encoding approaches. Concurrently, the \textbf{N}etwork-Specific \textbf{A}ugmentation (\textbf{NA}) techniques, detailed in Subsections~\ref{sec:system:training:off_aug} \& ~\ref{sec:system:training:on_aug}, enhance data diversity and further support robust generalization. The synergistic combination of time-aware representation learning and augmentation-driven robustness constitutes the core of the proposed early intrusion detection system, \textbf{\athena}.

\subsection{Problem Definition}

As defined in Equations~\ref{eq:flow_def} and~\ref{eq:timestamps}, a flow is characterized by its packet sequence $F$ and corresponding timestamp sequence $T$. The model therefore processes the combined packet--timestamp pairs
\begin{equation*}
    \{(p_i, t_i) \mid i = 0, 1, \dots, n{-}1\},
\end{equation*}
where $p_i \in \mathbb{R}^d$ denotes the $i$-th packet and $t_i$ its arrival time.

The goal of early intrusion detection is to learn a parameterized function $f_{\theta}$ that predicts the class label $y \in \mathcal{Y}$, where $\mathcal{Y}$ contains the benign and attack classes of interest, using only a prefix of the flow. At decision step $k$, with $1 \leq k \leq \min(n, N)$ and $N$ denoting the model's maximum input length, the prediction is given by
\begin{equation*}
    \hat{\mathbf{y}}_k = f_{\theta}(F_{0:k-1}, T_{0:k-1}),
\end{equation*}
where $\hat{\mathbf{y}}_k$ is the predicted probability distribution over $\mathcal{Y}$. The objective is to correctly infer the true class $y$ while minimizing both the classification error and the required observation length $k$, subject to achieving a confidence level threshold $\tau$, which specifies the minimum required prediction confidence (see Section~\ref{sec:exp_meth:eval_metrics:conf_based}).

%% file: sections/4_implementation.tex
The implementation of \athena is structured into three main components: network data processing, model development, and deployment on edge devices. We employ Scapy~\cite{scapy}, a Python library for packet manipulation and network traffic analysis, to extract relevant packet flows from raw PCAP files, enabling efficient flow reconstruction and preprocessing. The Transformer-based detection models are developed using TensorFlow~\cite{tensorflow2016osdi}, incorporating time-aware positional encodings to capture fine-grained temporal dependencies within network flows, thereby improving early threat detection. To facilitate real-time inference on resource-constrained IoT devices, the trained models are deployed on the Raspberry Pi Zero 2 W using LiteRT, a lightweight runtime optimized for efficient deep learning execution. This deployment strategy ensures low-latency processing, high detection accuracy, and minimal resource overhead, making the system well-suited for practical IoT security applications where computational efficiency and rapid response times are critical. To support reproducibility and facilitate further research, the complete source code of \athena is publicly available on GitHub\footnote{\url{https://github.com/ioannispan/A-THENA}}.

%% file: sections/5_exp_methodology.tex
This section provides a detailed overview of the methodology employed for conducting experiments and systematically evaluating the performance of our system. It details the experimental setup, dataset characteristics, and evaluation framework, ensuring a rigorous and reproducible analysis of the proposed approach.

\subsection{Datasets}

To ensure a comprehensive and robust evaluation, this work significantly expands upon our prior work~\cite{splitech_paper}, which utilized a single dataset. We employ three publicly available benchmark datasets for intrusion detection in IoT environments: \mbox{CICIoT2023}~\cite{neto23sensors}, \mqtt~\cite{hindy21inc}, and \iotid~\cite{ullah20ai}. As outlined in Table~\ref{tab:attacks}, each dataset encompasses distinct attack categories. In the case of \mbox{CICIoT2023}, we specifically focus on web-based attacks, resulting in a subset of the dataset referred to as \ciciot. Consequently, incorporating the benign traffic class, the final classification tasks consist of 6, 5, and 9 classes for \ciciot, \mqtt, and \iotid, respectively.

\begin{table}[t]
    \small
    \caption{Attack Categories in the Selected Datasets}
    \label{tab:attacks}
    \begin{tabular}{l l}
        \toprule
        \textbf{\ciciot} & \textbf{\mqtt} \\
        \midrule
        (1) SQL Injection & \multirow{2}{*}{(1) Aggressive Scan} \\
        (2) Command Injection & \multirow{2}{*}{(2) UDP Scan} \\
        (3) Backdoor Malware & \multirow{2}{*}{(3) MQTT Brute-Force} \\
        (4) Uploading Attack & \multirow{2}{*}{(4) Sparta SSH Brute-Force} \\
        (5) Cross-Site Scripting (XSS) &  \\
        \midrule
        \multicolumn{2}{c}{\textbf{\iotid}} \\
        \midrule
        (1) DoS SYN Flooding & (5) Mirai UDP Flooding \\
        (2) Mirai ACK Flooding & (6) MiTM ARP Spoofing \\
        (3) Mirai Host Brute-Force & (7) Host \& Port Scan \\
        (4) Mirai HTTP Flooding & (8) OS Scan \\
        \bottomrule
    \end{tabular}
\end{table}

The selected datasets provide a diverse and representative set of IoT intrusion scenarios, enabling a comprehensive evaluation of the system's generalization capabilities. \ciciot focuses on web-based application-layer attacks (\textit{e.g.}, SQL Injection, Command Injection, XSS) that target IoT web services. \mqtt covers protocol-specific and reconnaissance attacks (\textit{e.g.}, MQTT Brute-Force, SSH Brute-Force, UDP Scans) affecting IoT communication protocols. \iotid includes Mirai botnet-driven DDoS attacks (\textit{e.g.}, SYN Flooding, HTTP Flooding, MiTM ARP Spoofing) representing large-scale IoT security threats. By testing across these datasets, the system is validated against application-layer, protocol-specific, and DDoS-based intrusions, demonstrating its robustness and real-world applicability in IoT security.

\begin{table}[t]
    \centering
    \small
    \caption{Dataset Statistics for \mqtt Before and After Offline Augmentation}
    \label{tab:mqtt_stats}
    \begin{tabular}{l | r r r r r | r }
        \toprule
        \multirow{2}{*}{\textbf{Split}} &
        \multirow{2}{*}{\textbf{Benign}} &
        \textbf{Aggressive} &
        \textbf{UDP} &
        \textbf{MQTT} &
        \textbf{Sparta SSH} & 
        \multirow{2}{*}{\textbf{Total}} \\
        &
        &
        \textbf{Scan} &
        \textbf{Scan} &
        \textbf{Brute-Force} &
        \textbf{Brute-Force} &
        \\
        \midrule
        \textbf{Original}   & 6015 & 2412 & 1322 & 6440 & 6440 & \textbf{22629} \\
        \midrule
        \textbf{Test}       & 602  & 241  & 132  & 644  & 644  & \textbf{2263} \\
        \midrule
        Training ($m_c$)    & 4331 & 1736 & 952  & 4636 & 4637 & 16292 \\
        $m_\mathrm{d}-m_c$  & $<$0 & 295  & 1079 & $<$0 & $<$0 & -- \\
        $a_c$               & 0.2  & 0.17 & 1.13 & 0.2  & 0.2  & -- \\
        Training ($m'_c$)   & 5211 & 2054 & 2034 & 5574 & 5562 & 20435 \\
        $z_c$               & 0.07 & 1.714 & 1.74 & 0    & 0.002 & -- \\
        \textbf{Training ($\boldsymbol{m''_c}$)}  & 5576 & 5575 & 5573 & 5574 & 5573 & \textbf{27871} \\
        \midrule
        Validation ($m_c$)  & 1082 & 435  & 238  & 1160 & 1159 & 4074 \\
        \textbf{Validation ($\boldsymbol{m'_c}$)} & 1322 & 512  & 514  & 1383 & 1391 & \textbf{5122} \\
        \bottomrule
    \end{tabular}
\end{table}

Table~\ref{tab:mqtt_stats} summarizes the offline augmentation pipeline for \mqtt, which comprises 22,629 samples across five classes. We first allocate 10\% of the data to the test set (2,263 samples), and then perform 5-fold cross-validation on the remaining 90\%, resulting in training and validation partitions of 16,292 and 4,074 samples, respectively. The training set exhibits moderate class imbalance, with the minority class (UDP Scan) containing 952 samples and the majority classes (MQTT Brute-Force, Sparta SSH Brute-Force) reaching 4,637 samples. Given the model's parameter count and number of classes, the target sample count per class is computed as $m_\mathrm{d} \approx (2P) / C = (2 \cdot 5{,}077) / 5 = 10{,}154 / 5 \approx 2{,}030.8 \to 2{,}031 $. Classes falling below this threshold (Aggressive Scan, UDP Scan) undergo augmentation proportional to their deficit, with UDP Scan receiving the highest augmentation factor ($a_2 = 1.13$) due to its minority status. Majority classes receive minimal augmentation ($a_0 = a_3 = a_4 = 0.2$) to inject early-stage detection patterns without exacerbating imbalance. Following subflow generation, the training set expands to 20,435 samples. Hybrid oversampling then equalizes all classes to $\sim$5,574 samples, yielding a final balanced training set of 27,871 samples. The validation set undergoes only subflow generation, expanding from 4,074 to 5,122 samples while preserving the original class distribution for unbiased evaluation.

The remaining datasets follow analogous augmentation procedures. \ciciot contains 2,240 samples across six classes, split into 1,611 training, 423 validation, and 226 test samples. Offline augmentation expands the training set to 10,175 balanced samples through intensive subflow generation (with augmentation factors up to 29 for underrepresented classes), while the validation set increases from 423 to 2,024 samples. \iotid comprises 19,612 samples across nine classes, divided into 14,120 training, 3,530 validation, and 1,962 test samples. Although not the largest dataset, it exhibits moderate class imbalance, which is mitigated through a combination of conservative augmentation for most classes and stronger augmentation for the smallest ones. Subflow generation expands the training set to 18,092 samples, and subsequent hybrid oversampling produces a balanced set of 31,140 samples; the validation set grows from 3,530 to 4,524 samples.

\subsection{Training Configuration}

Training is conducted using the Adam optimizer~\cite{adam2014iclr} and the EDL loss function (Subsection~\ref{sec:system:training:loss_fn}), with early stopping applied to the validation loss using a patience of 7 epochs. Both encoding selection and hyperparameter tuning are performed via 5-fold cross-validation. After identifying the best temporal encoding variant, the final model is trained on the full development set, using a 90\%/10\% split for training and validation. Dataset-specific hyperparameters are as follows: \ciciot uses a batch size of 4 and a learning rate of $10^{-4}$; \iotid uses a batch size of 8 and a learning rate of $10^{-4}$; and \mqtt uses a batch size of 8 and a learning rate of $2\times 10^{-4}$.

\subsection{Evaluation Metrics}
\label{sec:exp_meth:eval_metrics}

To assess the effectiveness of \athena, we design an evaluation process to closely replicate real-world deployment conditions. This approach ensures that performance metrics accurately reflect the system's practicality and reliability in real-world applications. In practice, when deployed on a host machine, the system continuously monitors network flows to promptly determine whether an attack is occurring. Given the critical need for early threat detection in cybersecurity, the evaluation focuses on both the accuracy and responsiveness of the system.

\subsubsection{Confidence-Based Performance Metrics}
\label{sec:exp_meth:eval_metrics:conf_based}

To evaluate performance, a set of confidence-based metrics is employed. Given a confidence threshold $\tau$ applied to the top-$1$ softmax score, the evaluation focuses on how quickly the system arrives at a confident decision. The process begins with the first packet of each test flow, gradually adding packets until the model's confidence reaches or exceeds $\tau$. If the threshold is not met after processing all $N$ packets, the final classification is based on the full sequence. Because $\tau$ directly dictates the trade-off between classification speed and accuracy, we treat it as a tunable hyperparameter; a detailed sensitivity analysis of its impact is provided in Section~\ref{sec:results:sensitivity}.

A key metric is Earliness (E), measuring the number of packets needed before the model reaches the confidence threshold and a correct prediction is made; across a batch or test set, we report the maximum earliness observed over all correctly classified flows, representing the worst-case detection delay. Lower values indicate faster, more efficient classification, which is crucial for real-time intrusion detection. At the threshold point, Top-$1$ Accuracy (A), \textit{i.e.}, the percentage of correctly classified flows, False Negative Rate (FNR), \textit{i.e.}, the proportion of attack flows that are incorrectly classified as benign, and False Alarm Rate (FAR), \textit{i.e.}, the proportion of benign flows mistakenly classified as attacks, are also calculated. A high FNR is particularly concerning, as undetected attacks pose a severe security risk. Respectively, minimizing FAR is essential to prevent unnecessary security alerts, which can lead to operational inefficiencies and desensitization to genuine threats.

Lastly, the Early Risk Detection Error (ERDE)~\cite{fernandez18nca} is utilized, a metric that evaluates both the correctness of the model's predictions and the delay in reaching a decision. ERDE is a parametric metric; flows requiring more than $o$ packets for accurate classification incur a higher penalty. Given a flow $F \in \mathcal{F}$, it is computed as follows:
\begin{displaymath}
    \mathrm{ERDE}_{o}(F) = 
    \begin{cases}
        \frac{TP}{|\mathcal{F}|} & \text{if } F \text{ is a False Positive} \\
        1 & \text{if } F \text{ is a False Negative} \\
        1 - \frac{1}{1 + \operatorname{exp}({t_\mathrm{d}-o})} & \text{if } F \text{ is a True Positive} \\
        0 & \text{if } F \text{ is a True Negative}
    \end{cases}
\end{displaymath}
where $TP$ denotes the number of True Positives, and $t_\mathrm{d}$ represents the number of packets required to correctly classify a malicious flow, \textit{i.e.}, the flow's earliness.

\subsubsection{Resource-Constrained Deployment Evaluation}

Beyond classification performance, the feasibility of deploying the proposed system on resource-limited edge devices is also assessed. To simulate real-world IoT applications, the system is deployed on the Raspberry Pi Zero $2$ W, a compact, low-power embedded device featuring a $1$GHz quad-core $64$-bit ARM Cortex A$53$ processor and $512$ MB of RAM, commonly used in IoT environments. This evaluation measures two key factors:
\begin{itemize}
    \item \textbf{Latency:} The time required for the model to process and classify a flow. Low latency is essential for real-time intrusion detection.
    \item \textbf{Memory Requirements:} The system's RAM and storage footprint, which determines whether it can be deployed on constrained IoT devices without excessive resource consumption.
\end{itemize}

The Raspberry Pi Zero $2$ W serves as an ideal test platform, as its limited processing power and memory reflect the constraints of real-world IoT deployments. By validating the system's efficiency in such an environment, its suitability for lightweight cybersecurity applications is demonstrated.

%% file: sections/6_results.tex
This section provides a detailed analysis of our evaluation results, highlighting the effectiveness of our approach in comparison to existing methods.

\subsection{Comparison Methods}

\begin{table}[t]
    \small
    \centering
    \begin{threeparttable}
    \caption{Summary of Positional Encoding Mechanisms}
    \label{tab:encoding_summary}
    \begin{tabular}{c | r | c r l}
    \toprule
    \textbf{Category} & \textbf{Encoding} & \textbf{Learnable} & \textbf{Parameter Count} & \textbf{Input Type} \\
    \midrule
    \multirow{6}{*}{\textbf{Non-}} & None & \xmark & 0 & --- \\
    \multirow{6}{*}{\textbf{Time-Aware}} & Embedding & \cmark & $d_m \cdot N = 240$ & discrete index \\
    & Convolutional & \cmark & $d_m \cdot (K \cdot d + 1) = 10760$ & discrete index \\
    & Global Relative & \cmark & $N \cdot d_h = 240$ & discrete index \\
    & Sinusoidal & \xmark & 0 & discrete index \\
    & Fourier & \cmark & $d_m/2 = 4$ & discrete index \\
    & RoPE & \xmark & 0 & discrete index \\
    \midrule
    \multirow{6}{*}{\textbf{\shortstack{Time-Aware \\ (Related Work)}}} & GTID~\cite{han23compsec} & \xmark & 0 & index + $\Delta t$ \\
    & FATA~\cite{zhang2023fata} & \cmark & 3 & index + timestamp \\
    & Time2Vec~\cite{miyamoto2024acsac} & \cmark & $2 \cdot d_m = 16$ & continuous timestamp \\
    & CTLPE~\cite{kim2024arxiv} & \cmark & $2 \cdot d_m = 16$ & continuous timestamp \\
    & Chrono~\cite{zhang2025arxiv}\tnote{a} & \cmark & $2 \cdot d_m = 16$ & timestamp + $\Delta t$ \\
    & PEA~\cite{morenocartagena2023icml} & \xmark & 0 & continuous timestamp \\
    \midrule
    \multirow{3}{*}{\textbf{\shortstack{Time-Aware\\(\athena)}}} & TA Sinusoidal & \xmark & 0 & continuous timestamp \\
    & TA Fourier & \cmark & $d_m/2 = 4$ & continuous timestamp \\
    & TA RoPE & \xmark & 0 & continuous timestamp \\
    \bottomrule
    \end{tabular}
    \begin{tablenotes}
    \small
    \item[a] Parameter count assumes an MLP with one fully-connected layer for relative embeddings; exact size may vary.
    \end{tablenotes}
    \end{threeparttable}
\end{table}

The proposed system is evaluated against a diverse range of methods, organized into four comparison groups. Table~\ref{tab:encoding_summary} provides a unified comparison of all evaluated encoding mechanisms, detailing their input types, learnability, and parameter complexity. Conceptual descriptions and novelty contrasts for each method are presented in Section~\ref{sec:back_related}; this subsection focuses on the experimental rationale for each group and any implementation-specific details required for reproducibility.

\subsubsection{Non-Time-Aware Positional Encodings}

We include seven traditional positional encoding mechanisms (top section of Table~\ref{tab:encoding_summary}) to quantify the benefit of temporal awareness by establishing a positional-information baseline. These methods---No Encoding, Embedding, Convolutional ($K{=}3$, $d_m$ filters), Global Relative~\cite{shaw18acl, huang19iclr}, Sinusoidal, Fourier, and RoPE---all operate on discrete indices and are detailed conceptually in Subsection~\ref{sec:back_related:encodings:trad}. Each encoding is integrated into the same base Transformer architecture (Section~\ref{base_model}), augmentation pipeline, and EDL training objective used by \athena, ensuring that observed differences are attributable solely to the positional encoding mechanism.

\subsubsection{Traditional Machine Learning Baselines}

To contextualize \athena's raw-traffic approach against the established feature-engineering paradigm (see Subsection~\ref{sec:back_related:ids:ml}), we compare against seven classical algorithms: Naive Bayes, Logistic Regression, kNN, SVM, MLP, Random Forest, and XGBoost. For these methods, we extract 77 statistical features from network flows using CICFlowMeter-V4.0~\cite{CICFlowMeter}, following standard practice in cybersecurity research~\cite{sarhan2021bdta, rodriguez2022sensors}. Features include packet count, statistical measures of packet lengths in both directions, TCP flag counts, and flow active time, with standard preprocessing (categorical-to-numerical conversion and standardization). Table~\ref{tab:ml_hyper} presents the hyperparameters selected for each algorithm and dataset following systematic tuning, balancing model complexity with classification performance.

\begin{table}[t]
    \small
    \centering
    \caption{Hyperparameters for Traditional Machine Learning Baselines}
    \label{tab:ml_hyper}
    \resizebox{\textwidth}{!}{
        \begin{tabular}{l r | l l l}
        \toprule
        \textbf{Algorithm} & \textbf{Hyperparameter} & \textbf{\ciciot} & \textbf{\mqtt} & \textbf{\iotid} \\
        \midrule
        \textbf{Naive Bayes} & - & - & - & - \\
        \midrule
        \textbf{Logistic Regression} & regularization ($C$) & 0.1 & 0.1 & 0.1 \\
        \textbf{(LR)} & max iterations & 2000 & 2000 & 2000 \\
        \midrule
        \multirow{2}{*}{\textbf{k-Nearest Neighbors}} & neighbors ($k$) & 7 & 3 & 9 \\
        \multirow{2}{*}{\textbf{(kNN)}} & metric & Euclidean & Euclidean & Euclidean \\
         & weights & distance & distance & distance \\
        \midrule
        \textbf{Support Vector Machine} & kernel & RBF & RBF & RBF \\
        \textbf{(SVM)} & regularization ($C$) & 10 & 0.1 & 10 \\
        \midrule
        \textbf{Multi-Layer Perceptron} & hidden layers & (64, 32) & (64, 32) & (64, 32) \\
        \textbf{(MLP)} & activation & ReLU & ReLU & ReLU \\
        \midrule
        \textbf{Random Forest} & estimators & 50 & 10 & 10 \\
        \textbf{(RF)}  & max depth & 10 & 10 & 10 \\
        \midrule
        \multirow{2}{*}{\textbf{XGBoost}} & estimators & 50 & 10 & 10 \\
         & eval metric & logloss & logloss & logloss \\
        \bottomrule
        \end{tabular}
    }
\end{table}

\subsubsection{Time-Aware Encodings from Related Work}

Six recent time-aware encodings (middle section of Table~\ref{tab:encoding_summary}) are included to benchmark \athena's temporal representation against the most relevant prior work. The conceptual motivation, mathematical formulations, and novelty contrasts for each method---GTID, FATA, Time2Vec, CTLPE, ChronoFormer, and PEA---are presented in Subsection~\ref{sec:back_related:encodings:time}. To ensure a fair comparison, we integrate each encoding into our base Transformer architecture (Section~\ref{base_model}), utilizing the identical augmentation pipeline and EDL objective. Specifically, we reimplement only the encoding component rather than the full architectures from the original studies, isolating the impact of temporal representation. For ChronoFormer's relative component, we employ the simplest possible MLP---one fully connected layer---that maps each $\Delta t$ to a vector of dimensionality $d_\textrm{m}$.

\subsubsection{Early Intrusion Detection Architectures}

We compare against four neural network architectures proposed for early intrusion detection~\cite{ahmad24icstw, islam23cloudcom, ahmad23icstw, ahmad22icstw}: eRNN, eTransformer, eAtt, and eGlo (see Subsection~\ref{sec:back_related:ids:early} for conceptual context). These models represent alternative architectural approaches to early detection but do not utilize packet timestamps, allowing us to assess both the architectural efficiency of \athena's lightweight design and the specific contribution of time-aware representation learning to early threat identification.

\subsection{Detection Performance Evaluation}

This subsection presents a comprehensive evaluation of \athena across three IoT intrusion detection datasets, assessing both its accuracy and its ability to provide timely predictions under the confidence-based early detection framework described in Section~\ref{sec:exp_meth:eval_metrics:conf_based}. For each model and baseline, we report Top-1 Accuracy (A, in percentage), Earliness (E, in packets), False Alarm Rate (FAR, in percentage), False Negative Rate (FNR, in percentage), and the Early Risk Detection Error computed on the first 5 packets (ERDE$_5$), all evaluated at a confidence threshold of $\tau = 0.95$, which we select based on the sensitivity analysis in Section~\ref{sec:results:sensitivity}, as it provides near-optimal accuracy while maintaining minimal impact on detection latency.

\subsubsection{Non-Time-Aware Encoding Baselines}

Table~\ref{tab:nta_encodings} presents a comparative performance analysis of traditional non-time-aware positional encodings versus the proposed time-aware variants across the three evaluated datasets. The results reveal distinct dataset-specific trends, highlighting the importance of evaluating intrusion detection systems on diverse network scenarios. Specifically, \ciciot shows the highest variability in performance across encoding methods, suggesting that encoding effectiveness depends heavily on the attack types present. In contrast, \mqtt demonstrates more stable performance, indicating reduced sensitivity to encoding variations, while \iotid highlights the importance of handling longer network flows, as its attack complexity necessitates extended packet sequences for accurate detection.

\begin{table}[t]
    \small
    \centering
    \caption{Performance of Non-Time-Aware Encoding Mechanisms}
    \label{tab:nta_encodings}
    \resizebox{\textwidth}{!}{
        \begin{tabular}{l|ccccc|ccccc|ccccc}
        \toprule
        \textbf{Encoding}
        & \multicolumn{5}{c|}{\textbf{\ciciot}} 
        & \multicolumn{5}{c|}{\textbf{\mqtt}} 
        & \multicolumn{5}{c}{\textbf{\iotid}} \\
        \textbf{Method} & \textbf{A} & \textbf{E} & \textbf{FAR} & \textbf{FNR} & \textbf{ERDE$_\textbf{5}$} 
        & \textbf{A} & \textbf{E} & \textbf{FAR} & \textbf{FNR} & \textbf{ERDE$_\textbf{5}$}
        & \textbf{A} & \textbf{E} & \textbf{FAR} & \textbf{FNR} & \textbf{ERDE$_\textbf{5}$} \\
        \midrule
        None             
        & 70.80 & \textbf{1} & 100.0 & \textbf{0.0} & 0.152 
        & 89.35 & 30 & 19.93 & \textbf{0.0} & 0.045
        & 80.28 & \textbf{30} & \textbf{0.0} & 1.41 & \textbf{0.124} \\
        Embedding        
        & 83.19 & 8 & 60.31 & \textbf{0.0} & 0.134 
        & 90.68 & 5 & 19.93 & \textbf{0.0} & 0.052
        & 85.37 & \textbf{30} & \textbf{0.0} & 2.56 & 0.257 \\
        Convolutional    
        & 79.20 & 2 & 100.0 & \textbf{0.0} & 0.157 
        & 92.00 & \textbf{1} & 20.10 & \textbf{0.0} & 0.046
        & 89.25 & \textbf{30} & \textbf{0.0} & \textbf{0.96} & 0.328 \\
        Global Relative  
        & 83.19 & 3 & 100.0 & \textbf{0.0} & 0.167 
        & 93.33 & 3 & 13.29 & \textbf{0.0} & 0.036
        & 84.66 & \textbf{30} & \textbf{0.0} & 1.60 & 0.179 \\
        \midrule
        Sinusoidal  
        & 86.73 & 2 & 20.63 & \textbf{0.0} & 0.044 
        & 92.58 & 4 & 14.78 & \textbf{0.0} & 0.042
        & 79.41 & \textbf{30} & \textbf{0.0} & 5.70 & 0.199 \\
        Fourier     
        & 88.94 & 2 & 49.21 & \textbf{0.0} & 0.086 
        & 95.01 & 5 & 12.46 & \textbf{0.0} & 0.046
        & 85.73 & \textbf{30} & \textbf{0.0} & 3.07 & 0.304 \\
        RoPE        
        & 77.88 & \textbf{1} & 100.0 & \textbf{0.0} & 0.153 
        & 93.99 & \textbf{1} & 13.29 & \textbf{0.0} & 0.035
        & 87.67 & \textbf{30} & \textbf{0.0} & 2.94 & 0.297 \\
        \midrule
        \textbf{TA Sinusoidal} 
        & \textbf{100.0} & \textbf{1} & \textbf{0.0} & \textbf{0.0} & \textbf{0.015} 
        & 96.69 & 3 & \textbf{0.0} & \textbf{0.0} & 0.017
        & 69.16 & \textbf{30} & \textbf{0.0} & 19.46 & 0.409 \\
        \textbf{TA Fourier}    
        & \textbf{100.0} & 4 & \textbf{0.0} & \textbf{0.0} & 0.059 
        & \textbf{100.0} & \textbf{1} & \textbf{0.0} & \textbf{0.0} & \textbf{0.014}
        & 87.67 & \textbf{30} & \textbf{0.0} & 1.41 & 0.245 \\
        \textbf{TA RoPE}       
        & 83.19 & 7 & \textbf{0.0} & \textbf{0.0} & 0.029 
        & \textbf{100.0} & 4 & \textbf{0.0} & \textbf{0.0} & 0.021
        & \textbf{93.83} & \textbf{30} & \textbf{0.0} & 1.34 & 0.422 \\
        \midrule
        \textbf{\athena} 
        & \textbf{100.0} & \textbf{1} & \textbf{0.0} & \textbf{0.0} & \textbf{0.015} 
        & \textbf{100.0} & \textbf{1} & \textbf{0.0} & \textbf{0.0} & \textbf{0.014}
        & \textbf{93.83} & \textbf{30} & \textbf{0.0} & 1.34 & 0.422 \\
        \bottomrule
        \end{tabular}
    }
\end{table}

\athena's time-aware hybrid encoding consistently yields the highest accuracy across all evaluated datasets, underscoring its effectiveness in modeling the temporal dynamics of network traffic. In contrast, traditional positional encodings perform notably worse, with \athena achieving accuracy gains of 11.06, 4.99, and 4.58 percentage points over the best non-time-aware alternatives on \ciciot, \mqtt, and \iotid, respectively. This performance gap can be attributed to the inability of conventional encodings to account for irregular packet inter-arrival times, which carry critical discriminative information for detecting anomalies in IoT traffic.

Among the non-time-aware baselines, models without any positional encoding perform particularly poorly on \ciciot (70.80\% accuracy), demonstrating that some form of positional information is essential. Learned positional encodings (Embedding and Convolutional) show mixed results, with the Convolutional approach achieving strong performance on \iotid (89.25\% accuracy) but struggling on \ciciot (79.20\%). The classical function-based encodings (Sinusoidal, Fourier, RoPE) generally outperform learned approaches, with Fourier encoding reaching 95.01\% accuracy on \mqtt, yet still falling short of time-aware variants. Notably, Global Relative encoding, which incorporates relative positional relationships, achieves competitive accuracy (93.33\% on \mqtt) but cannot match the temporal precision provided by time-aware methods. Moreover, FAR values drop to 0\% across all datasets when using time-aware variants, eliminating unnecessary alerts and stabilizing early prediction behavior. While ERDE$_5$ scores do not always align directly with accuracy, this discrepancy primarily reflects differences in earliness rather than model uncertainty, consistent with observations that marginal variations in earliness have limited operational impact on end-to-end latency (see Subsection~\ref{dl_latency_mf}).

The final model for \athena for each dataset is selected based on the lowest validation loss, as reported in Table~\ref{tab:cv_results}. Each dataset converges to a different time-aware encoding variant---sinusoidal for \ciciot, Fourier for \mqtt, and RoPE for \iotid---further confirming the importance of \athena's hybrid encoding design. Rather than relying on a single temporal representation, \athena flexibly adapts to the temporal characteristics of each dataset, enabling optimal performance across diverse IoT traffic patterns. The distinct selection of encodings demonstrates that no single temporal mechanism is universally optimal; instead, the hybrid design allows the model to exploit dataset-specific temporal structures, ultimately yielding the strongest accuracy, FAR/FNR, and earliness trade-offs observed in our experiments.

\begin{table}[t]
    \centering
    \small
    \caption{Five-Fold Cross-Validation Results Across Datasets}
    \label{tab:cv_results}
    \resizebox{\textwidth}{!}{
        \begin{tabular}{ll|ccc}
        \toprule
        \multicolumn{2}{l|}{\textbf{Metric}} 
        & \textbf{\ciciot} 
        & \textbf{\mqtt} 
        & \textbf{\iotid} \\
        \midrule
        \multirow{2}{*}{\textbf{Validation}} 
        & \textbf{TA Sinusoidal} 
        & $\mathbf{2.24 \times 10^{-7} \pm 4.1 \times 10^{-9}}$ 
        & $1.80 \times 10^{-6} \pm 3.2 \times 10^{-8}$ 
        & $7.37 \times 10^{-2} \pm 1.9 \times 10^{-3}$ \\
        \multirow{2}{*}{\textbf{Loss}} 
        & \textbf{TA Fourier}
        & $3.85 \times 10^{-7} \pm 5.6 \times 10^{-9}$ 
        & $\mathbf{9.85 \times 10^{-8} \pm 1.4 \times 10^{-9}}$ 
        & $6.40 \times 10^{-2} \pm 1.6 \times 10^{-3}$ \\
        & \textbf{TA RoPE}
        & $3.69 \times 10^{-6} \pm 7.5 \times 10^{-8}$ 
        & $1.03 \times 10^{-6} \pm 2.1 \times 10^{-8}$ 
        & $\mathbf{5.48 \times 10^{-2} \pm 1.3 \times 10^{-3}}$ \\
        \midrule
        \multicolumn{2}{l|}{\textbf{Accuracy}} 
        & $99.87 \pm 0.05$ 
        & $100.0 \pm 0.00$ 
        & $94.41 \pm 0.22$ \\
        \multicolumn{2}{l|}{\textbf{Earliness}} 
        & $1.2 \pm 0.4$
        & $1.6 \pm 0.8$
        & $30.0 \pm 0.0$ \\
        \multicolumn{2}{l|}{\textbf{FAR}} 
        & $0.08 \pm 0.12$ 
        & $0.00 \pm 0.00$
        & $0.11 \pm 0.10$ \\
        \multicolumn{2}{l|}{\textbf{FNR}} 
        & $0.03 \pm 0.05$ 
        & $0.00 \pm 0.00$
        & $1.62 \pm 0.21$ \\
        \multicolumn{2}{l|}{\textbf{ERDE$_5$}} 
        & $0.017 \pm 0.005$ 
        & $0.013 \pm 0.003$
        & $0.488 \pm 0.012$ \\
        \bottomrule
        \end{tabular}
    }
\end{table}

\paragraph{Interpretation of Perfect Accuracy Results}
Careful and nuanced interpretation is required by the perfect test-set accuracy reported for CICIoT23-WEB and MQTT-IoT-IDS2020. This outcome is driven by several factors: (1) the attacks in these datasets exhibit strong temporal regularities that are effectively captured by the time-aware encodings; (2) the distinction between benign and malicious flows is substantial, leading to limited intra-class variability; and (3) the combined use of raw packet payloads and temporal information yields highly discriminative flow representations. To verify that such results are not artifacts of overfitting or data leakage, Table~\ref{tab:cv_results} reports the corresponding 5-fold cross-validation outcomes. As shown, the cross-validation metrics are closely aligned with the final test-set performance and exhibit extremely low variance across folds, supporting that the perfect accuracy achieved on these datasets reflects genuine model robustness rather than memorization.

\paragraph{Interpretability of Early Predictions (E = 1)}

Our early detection analysis reveals a counterintuitive result: \athena achieves perfect accuracy at E = 1 on \ciciot and \mqtt, where only the first packet is observed. Since the first packet’s timestamp is normalized to $t_0 = 0$, both time-aware and non-time-aware models receive identical temporal input at inference, meaning time cannot serve as a direct discriminative feature at this step. The performance gap at E = 1 therefore reflects a powerful training-time inductive bias rather than an inference-time artifact. During training, the model processes full sequence lengths up to $N=30$, allowing self-attention to propagate temporal information from later packets back to earlier representations. This is analogous to how a language model trained on full sentences learns richer, more contextualized single-word representations than one trained on isolated tokens; the temporal context seen during training shapes the embedding space even when that context is absent at inference.

\begin{figure*}[t]
    \centering
    \includegraphics[width=\linewidth]{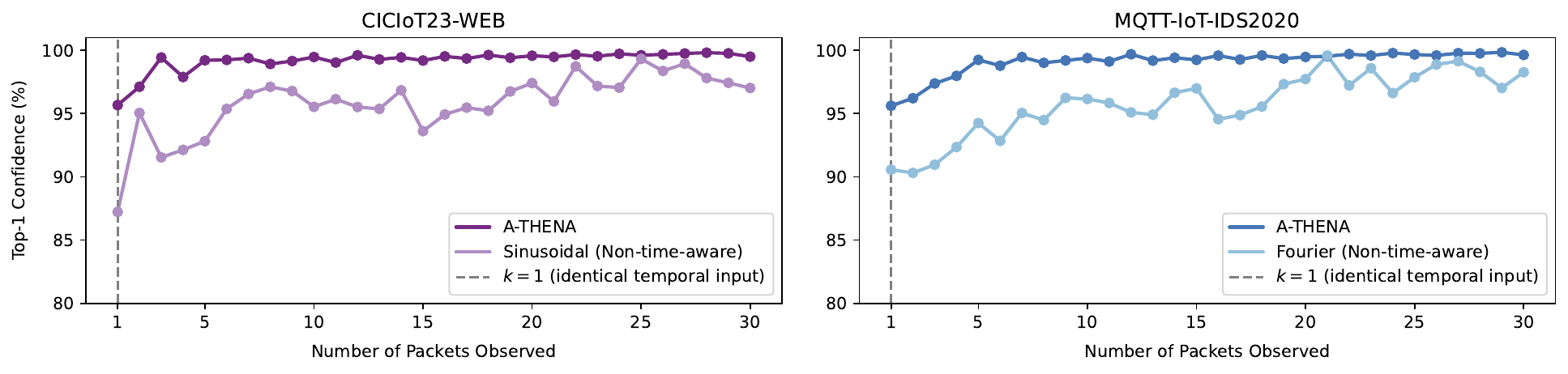}
    \caption{Prediction confidence trajectories as the number of observed packets ($k$) increases. The plots contrast \athena against corresponding non-time-aware baselines for a representative \textit{Backdoor Malware} attack (left) and an \textit{MQTT Brute-Force} attack (right).}
    \Description{Two line graphs showing top confidence percentages on the y-axis against observation length (k) on the x-axis. The left graph for CICIoT23-WEB compares A-THENA against a Sinusoidal baseline for a Backdoor Malware attack. A-THENA starts at roughly 95.5 percent confidence at k=1 and quickly stabilizes near 100 percent, while the Sinusoidal baseline starts much lower at 87 percent and fluctuates before slowly rising. The right graph for MQTT-IoT-IDS2020 compares A-THENA against a Fourier baseline for an MQTT Brute-Force attack. A-THENA begins at 95.5 percent at k=1 and smoothly approaches 100 percent, whereas the Fourier baseline starts at 91 percent and requires over 20 packets to reach 97 percent confidence.}
    \label{fig:confidence_trajectories}
\end{figure*}

Figure~\ref{fig:confidence_trajectories} provides direct visual evidence of this effect. It shows the top-1 confidence trajectory as a function of packets observed ($k$) for a representative Backdoor Malware flow from \ciciot and an MQTT Brute-Force flow from \mqtt, comparing \athena against the best-performing non-time-aware baseline on each dataset. At $k=1$, \athena already exhibits markedly higher confidence than the non-time-aware baseline on both datasets, despite both models receiving identical input at that step. Furthermore, \athena's confidence curve is notably more stable across subsequent packets, reflecting the more discriminative and temporally coherent representations learned through time-aware training. These observations confirm that time-aware encodings improve performance primarily through better learned representations rather than direct exploitation of timestamp values at inference.

\subsubsection{Comparison with Traditional Machine Learning}

\begin{table*}[t]
    \small 
    \centering
    \caption{Comparison With Feature-Based Methods}
    \label{tab:ml_models}
    \resizebox{\textwidth}{!}{
        \begin{tabular}{l|ccccc|ccccc|ccccc}
        \toprule
        \textbf{Encoding}
        & \multicolumn{5}{c|}{\textbf{\ciciot}} 
        & \multicolumn{5}{c|}{\textbf{\mqtt}} 
        & \multicolumn{5}{c}{\textbf{\iotid}} \\
        \textbf{Method} & \textbf{A} & \textbf{E} & \textbf{FAR} & \textbf{FNR} & \textbf{ERDE$_\textbf{5}$} 
        & \textbf{A} & \textbf{E} & \textbf{FAR} & \textbf{FNR} & \textbf{ERDE$_\textbf{5}$}
        & \textbf{A} & \textbf{E} & \textbf{FAR} & \textbf{FNR} & \textbf{ERDE$_\textbf{5}$} \\
        \midrule
        Naive Bayes
        & 43.36 & \textbf{1} & \textbf{0.0} & \textbf{0.0} & \textbf{0.006} 
        & 92.80 & \textbf{1} & 3.99 & 3.49 & 0.048 
        & 40.72 & \textbf{1} & 93.25 & 4.35 & \textbf{0.135} \\
        LR       
        & 75.22 & 30 & 42.86 & \textbf{0.0} & 0.616 
        & 97.61 & 29 & 7.97 & \textbf{0.0} & 0.580 
        & 86.90 & 30 & 16.75 & 1.79 & 0.715 \\
        kNN
        & 80.09 & 30 & 46.03 & \textbf{0.0} & 0.429 
        & 93.19 & 12 & 8.14 & \textbf{0.0} & 0.151
        & 68.65 & 30 & 69.00 & 1.41 & 0.299 \\
        SVM 
        & 81.86 & 30 & 33.33 & \textbf{0.0} & 0.612 
        & 96.86 & 30 & 11.96 & 1.02 & 0.448 
        & 89.65 & 30 & 10.00 & \textbf{0.42} & 0.655 \\
        MLP
        & 73.89 & 30 & 46.03 & 0.61 & 0.674 
        & 97.57 & 30 & 7.97 & \textbf{0.0} & 0.509 
        & 88.33 & 30 & 11.75 & 1.66 & 0.581 \\
        Random Forest
        & 93.81 & 30 & \textbf{0.0} & \textbf{0.0} & 0.666 
        & 98.85 & 30 & 1.99 & \textbf{0.0} & 0.140 
        & 90.11 & 30 & 1.25 & 6.66 & 0.593 \\
        XGBoost      
        & 82.74 & 30 & \textbf{0.0} & 12.27 & 0.265 
        & 98.01 & 30 & 9.80 & \textbf{0.0} & 0.759 
        & 89.25 & 30 & 6.75 & 1.54 & 0.806 \\
        \midrule
        \textbf{\athena} 
        & \textbf{100.0} & \textbf{1} & \textbf{0.0} & \textbf{0.0} & 0.015 
        & \textbf{100.0} & \textbf{1} & \textbf{0.0} & \textbf{0.0} & \textbf{0.014}
        & \textbf{93.83} & 30 & \textbf{0.0} & 1.34 & 0.422 \\
        \bottomrule
        \end{tabular}
    }
\end{table*}

Table~\ref{tab:ml_models} compares \athena against seven classical machine learning algorithms operating on hand-crafted statistical features. The results demonstrate \athena's substantial superiority over all traditional machine learning baselines across the three datasets. On \ciciot, \athena achieves perfect 100\% accuracy, outperforming the best traditional method (Random Forest at 93.81\%) by 6.19 percentage points. The performance gap widens dramatically for simpler classifiers: Naive Bayes manages only 43.36\% accuracy, while Logistic Regression and MLP achieve 75.22\% and 73.89\%, respectively. On \mqtt, \athena again reaches perfect accuracy, surpassing the strongest baseline (Random Forest at 98.85\%) by 1.15 points. For \iotid, \athena achieves 93.83\% accuracy, exceeding the best traditional method (Random Forest at 90.11\%) by 3.72 percentage points.

Among the traditional approaches, ensemble methods consistently outperform individual classifiers. Random Forest emerges as the strongest baseline across all datasets, achieving competitive accuracy and maintaining low false alarm rates on \ciciot (0\% FAR) and \mqtt (1.99\% FAR). However, it struggles with \iotid's complex attacks, producing a notably high 6.66\% FNR---meaning approximately one in fifteen attacks would go undetected. XGBoost demonstrates comparable accuracy but exhibits inconsistent false detection rates, including an alarming 12.27\% FNR on \ciciot and elevated FAR values across datasets.

A crucial observation concerns the earliness metric: almost all traditional ML methods require complete flows (30 packets) for confident classification, as feature extraction necessitates computing statistical aggregates over entire sessions. In stark contrast, \athena achieves confident predictions from just a single packet on \ciciot and \mqtt. This fundamental architectural difference---processing raw packets sequentially versus extracting session-level statistics---enables \athena to identify threats at the earliest possible moment, a capability that feature-based approaches cannot match.

\subsubsection{Time-Aware Encoding Mechanisms}

\begin{table}[t]
    \small
    \centering
    \caption{Comparison With Related Work}
    \label{tab:ta_encodings}
    \resizebox{\textwidth}{!}{
        \begin{tabular}{l|ccccc|ccccc|ccccc}
        \toprule
        \textbf{Encoding}
        & \multicolumn{5}{c|}{\textbf{\ciciot}} 
        & \multicolumn{5}{c|}{\textbf{\mqtt}} 
        & \multicolumn{5}{c}{\textbf{\iotid}} \\
        \textbf{Method} & \textbf{A} & \textbf{E} & \textbf{FAR} & \textbf{FNR} & \textbf{ERDE$_\textbf{5}$} 
        & \textbf{A} & \textbf{E} & \textbf{FAR} & \textbf{FNR} & \textbf{ERDE$_\textbf{5}$}
        & \textbf{A} & \textbf{E} & \textbf{FAR} & \textbf{FNR} & \textbf{ERDE$_\textbf{5}$} \\
        \midrule
        GTID~\cite{han23compsec}             
        & 87.17 & 7 & \textbf{0.0} & \textbf{0.0} & 0.031
        & 92.84 & 30 & 13.95 & 1.02 & 0.152 
        & 78.64 & 30 & 12.25 & 5.25 & 0.139 \\
        FATA~\cite{zhang2023fata}        
        & 83.19 & 5 & \textbf{0.0} & 20.25 & 0.054 
        & 95.01 & 5 & 9.97 & \textbf{0.0} & 0.057 
        & 88.33 & 30 & \textbf{0.0} & \textbf{0.0} & 0.103 \\
        Time2Vec~\cite{miyamoto2024acsac}    
        & 88.94 & 5 & 66.67 & \textbf{0.0} & 0.134
        & 97.48 & 30 & 10.13 & \textbf{0.0} & 0.149 
        & 88.89 & 30 & 1.75 & 1.47 & 0.084 \\
        CTLPE~\cite{kim2024arxiv} 
        & 80.97 & 24 & 57.14 & 3.07 & 0.136 
        & 94.30 & 2 & 14.29 & \textbf{0.0} & 0.037 
        & 84.45 & 30 & 3.25 & 0.45 & 0.078 \\
        Chrono~\cite{zhang2025arxiv}
        & 83.19 & 11 & 33.33 & 13.50 & 0.185
        & 96.82 & 30 & 7.97 & \textbf{0.0} & 0.191 
        & 75.94 & 30 & \textbf{0.0} & 2.50 & 0.200 \\
        PEA~\cite{morenocartagena2023icml}
        & 72.12 & 30 & 82.54 & \textbf{0.0} & 0.138 
        & 92.89 & 30 & 15.61 & \textbf{0.0} & 0.086 
        & 78.24 & 30 & \textbf{0.0} & 1.47 & 0.150  \\
        \midrule
        eRNN~\cite{ahmad24icstw}       
        & 86.73 & 17 & 19.05 & \textbf{0.0} & 0.090 
        & 94.83 & 6 & 18.44 & \textbf{0.0} & 0.058 
        & 91.34 & 30 & \textbf{0.0} & \textbf{0.0} & 0.159 \\
        eTransformer~\cite{islam23cloudcom}      
        & 91.59 & 30 & \textbf{0.0} & \textbf{0.0} & 0.111 
        & 95.58 & 3 & 11.13 & \textbf{0.0} & 0.033 
        & 85.58 & 30 & \textbf{0.0} & 0.45 & 0.094 \\
        eAtt~\cite{ahmad23icstw}
        & 89.82 & 3 & 19.05 & \textbf{0.0} & 0.047 
        & 95.23 & \textbf{1} & 19.10 & \textbf{0.0} & 0.045 
        & 89.30 & 30 & \textbf{0.0} & \textbf{0.0} & 0.181 \\
        eGlo~\cite{ahmad22icstw}
        & 88.05 & 2 & 28.57 & \textbf{0.0} & 0.054 
        & 89.35 & \textbf{1} & 33.39 & \textbf{0.0} & 0.067 
        & 82.31 & \textbf{3} & \textbf{0.0} & 4.16 & \textbf{0.053} \\
        \midrule
        \textbf{\athena} 
        & \textbf{100.0} & \textbf{1} & \textbf{0.0} & \textbf{0.0} & \textbf{0.015} 
        & \textbf{100.0} & \textbf{1} & \textbf{0.0} & \textbf{0.0} & \textbf{0.014}
        & \textbf{93.83} & 30 & \textbf{0.0} & 1.34 & 0.422 \\
        \bottomrule
        \end{tabular}
    }
\end{table}

Table~\ref{tab:ta_encodings} (top section) summarizes the performance of the six time-aware positional encoding mechanisms drawn from related work. Across all datasets, \athena's hybrid encoding delivers the highest overall effectiveness. Among the competing approaches, Time2Vec stands out as the strongest baseline. On \ciciot, \athena outperforms Time2Vec by 11.06 percentage points while avoiding the latter's exceptionally high false alarm rate of 66.67\%. On \mqtt, \athena achieves perfect accuracy with zero false alarms, surpassing Time2Vec's accuracy of 97.48\%. On \iotid, it further improves upon Time2Vec by an additional 4.94 percentage points.

Among related methods, performance varies substantially. GTID and FATA show strengths on individual datasets but suffer from high false detection rates or inconsistent behavior across benchmarks. Time2Vec achieves competitive accuracy but consistently exhibits elevated FAR values that limit practical usability. CTLPE and Chrono demonstrate moderate performance but struggle with inconsistent earliness or degraded accuracy on at least one dataset. PEA, which injects temporal information at the output rather than the input or attention layers, underperforms across all benchmarks. Overall, related encodings often trade accuracy for high false detection rates or unstable earliness. In contrast, \athena uniquely combines high accuracy with 0\% FAR on all datasets, demonstrating that integrating time-awareness directly into the input representation and attention computation provides the most robust and reliable temporal modeling for IoT intrusion detection.

\subsubsection{Comparison with Early Detection DL Architectures}

Table~\ref{tab:ta_encodings} (bottom section) reports the performance results for four specialized neural network architectures explicitly designed for early intrusion detection: eRNN, eTransformer, eAtt, and eGlo. Critically, these models do not incorporate packet timestamps in their classification process, relying solely on packet content and sequence order. This comparison therefore isolates the contribution of temporal awareness versus purely architectural innovations for early threat identification.

\athena demonstrates substantially stronger performance across all datasets, avoiding the critical trade-offs observed in competing architectures. On \ciciot, it achieves an 8.41 percentage point improvement over the best alternative, eTransformer (91.59\%), while maintaining zero false alarms. On \mqtt, \athena surpasses the strongest baseline by 4.42 percentage points; in contrast, other fast-response models like eAtt and eGlo incur unacceptable FAR values of up to 33.39\% to match \athena's single-packet earliness. Similarly, on \iotid, \athena secures a 2.49 percentage point gain over the closest competitor (eRNN), ensuring superior overall detection even as baselines like eAtt and eGlo sacrifice significant detection power to achieve comparable stability or speed.

These results validate \athena's core hypothesis: incorporating temporal information through time-aware encodings provides advantages that architectural innovations alone cannot replicate. While early detection architectures demonstrate that specialized designs can improve performance over general-purpose models, they fundamentally lack the temporal discriminability that timing patterns provide. By unifying a lightweight Transformer architecture with adaptive time-aware encodings, \athena achieves superior accuracy, earliness, and false detection rates---establishing a new standard for IoT intrusion detection systems.

\subsection{Computational Efficiency and Deployment Feasibility}

\begin{table}[t]
    \small
    \caption{Model Complexity and Efficiency Comparison ($n = 30$)}
    \label{tab:latency_mf}
    \begin{tabular}{r | r r r r}
    \toprule
    \multirow{3}{*}{\textbf{Model}} & \multirow{2}{*}{\textbf{Size on}} & \textbf{Average} & \textbf{Average} & \multirow{2}{*}{\textbf{Memory}} \\
    & \multirow{2}{*}{\textbf{Disk}} & \textbf{Inference} & \textbf{End-to-End} & \multirow{2}{*}{\textbf{Footprint}} \\
    & & \textbf{Latency} & \textbf{Latency} & \\
    \midrule
    Naive Bayes & 7.0 - 12.0 KB & 2.48 ms & 666.14 ms & \textbf{4.98 MB} \\
    LR & 4.0 - 6.5 KB & \textbf{0.84 ms} & 664.50 ms & 12.36 MB \\
    kNN & 6.34 - 19.43 MB & 7.69 ms & 671.35 ms & 17.63 MB \\
    SVM & 2.84 - 8.86 MB & \textbf{2.00 ms} & 665.66 ms & 12.76 MB \\
    MLP & 180.0 - 183.1 KB & \textbf{1.54 ms} & 665.20 ms & 9.62 MB \\
    Random Forest & 163.1 - 793.2 KB & 25.20 ms & 688.86 ms & 20.30 MB \\
    XGBoost & 51.6 - 244.1 KB & 3.66 ms & 667.32 ms & 17.74 MB \\
    \midrule
    eRNN~\cite{ahmad24icstw} & \textasciitilde 211.5 KB & 20.12 ms & 20.12 ms & 9.13 MB \\
    eTransformer~\cite{islam23cloudcom} & \textasciitilde 4.9 MB & 35.79 ms & 35.79 ms & 7.96 MB \\
    eAtt~\cite{ahmad23icstw} & \textasciitilde 68.5 KB & \textbf{1.39 ms} & \textbf{1.39 ms} & \textbf{3.38 MB} \\
    eGlo~\cite{ahmad22icstw} & \textasciitilde 74.3 KB & \textbf{0.57 ms} & \textbf{0.57 ms} & \textbf{3.25 MB} \\
    \midrule
    \textbf{\athena} & \textbf{\textasciitilde 40.0 KB} & \textbf{1.42 ms} & \textbf{1.42 ms} & \textbf{3.25 MB} \\
    \bottomrule
    \end{tabular}
\end{table}

Beyond detection accuracy, the practical viability of an intrusion detection system in IoT environments depends heavily on its computational footprint and real-time responsiveness. Edge devices typically operate under strict constraints on model size, memory availability, and processing latency. This subsection evaluates \athena's efficiency along these dimensions on a Raspberry Pi Zero~2~W, benchmarking it against both traditional feature-based machine learning pipelines and state-of-the-art deep learning architectures.

Table~\ref{tab:latency_mf} presents the storage requirements, latency, and runtime memory footprint for all models when classifying flows of 30 packets. Average latency values of $\leq$2\,ms and memory footprints of $\leq$5\,MB are shown in bold. Inference latency measurements correspond to the mean over 1,000 runs per model, while memory footprints include the overhead of the Python interpreter for LiteRT models and the \texttt{scikit-learn} execution environment for traditional ML baselines. Consequently, all memory values should be interpreted as approximate but reflective of realistic deployment conditions.

\subsubsection{Overhead of Traditional Feature Extraction Pipelines}

Table~\ref{tab:latency_mf} highlights the dominant role of feature extraction in the computational overhead of traditional machine learning pipelines. Although feature-based classifiers exhibit fast inference once features are available---requiring only 0.84~ms for Logistic Regression and 2.00~ms for SVM---the CICFlowMeter preprocessing stage introduces an average delay of 664~ms for a 30-packet flow. This preprocessing cost overwhelmingly determines the end-to-end latency for all classical methods. Crucially, this 664~ms delay represents an inherent architectural limitation rather than an inefficiency of the classifiers themselves. Computing the 77 statistical features requires multiple passes over the full packet sequence, maintaining internal state, and performing aggregate computations. Even at the smallest scale, extracting features for single-packet flows still requires 290~ms on average, underscoring the intrinsic expense of the feature computation process.

Memory usage further complicates deployment of traditional pipelines. Although feature extraction alone consumes a modest 3.45~MB, the total runtime memory required to compute features, load and execute the classifier reaches up to 20.30~MB for the most accurate model (Random Forest)---more than six times higher than \athena's footprint. Storage requirements also vary significantly across datasets. Compact linear models remain small (4.0--6.5~KB for Logistic Regression and 7.0--12.0~KB for Naive Bayes), but instance-based methods scale poorly: kNN ranges from 6.34 to 19.43~MB depending on training set size, while SVM models range from 2.84 to 8.86~MB. This variability creates uncertainty for deployment planning, as storage must accommodate worst-case model sizes.

Finally, the feature extraction paradigm fundamentally conflicts with the objective of early detection. Because statistical features are meaningful only after a sufficient number of packets have been observed, nearly all traditional classifiers require full 30-packet flows (E = 30 in Table~\ref{tab:ml_models}) before producing a confident prediction. Metrics such as average packet length or directional traffic ratios cannot be inferred from the first few packets and as a result, even a perfectly accurate classical classifier would still incur the 664~ms preprocessing delay and the requirement for complete flows, yielding detection latencies that are unsuitable for real-time IoT threat response.

\subsubsection{Latency and Memory Footprint of Early Detection DL Architectures}
\label{dl_latency_mf}

Table~\ref{tab:latency_mf} confirms that operating directly on raw packets eliminates the feature-extraction bottleneck. Because \athena and competing deep learning architectures ingest packet sequences directly, their end-to-end latency effectively converges with their inference time. This architectural advantage allows \athena to achieve detection speeds approximately $480\times$ faster than feature-based pipelines, delivering a classification verdict before traditional methods have even finished computing the necessary flow statistics.

\athena achieves a favorable combination of inference speed, compactness, and memory efficiency. With a latency of just 1.42~ms for 30-packet flows, the system supports near real-time threat detection with minimal operational delay. Its 40~KB storage size is the smallest among all evaluated deep learning models and substantially smaller than the storage demands of feature-based ensemble models. The 3.25~MB runtime memory footprint matches eGlo as the most memory-efficient solution, making \athena suitable for deployment on devices with tight memory budgets. Experimental measurements confirm that the system's latency scales sub-linearly with sequence length: a single-packet flow requires only 0.17~ms, indicating that input length has minimal impact on runtime. Latency remains consistent across all positional encodings tested, with differences below 0.05~ms---within expected measurement variation. Instead, latency is largely determined by model size, underscoring the importance of \athena's compact architectural design.

Compared to other early-detection architectures, \athena provides a superior balance of accuracy, speed, and resource usage. While eGlo and eAtt offer fast inference, their representational capacity is insufficient to achieve high accuracy across all datasets. Conversely, larger models such as eTransformer deliver greater capacity at the cost of significantly higher latency and memory overhead. \athena's hybrid temporal encoding and lightweight architecture allow it to maintain low latency without compromising detection performance.

These findings establish \athena as well-suited for resource-constrained IoT environments. By removing the feature extraction stage, maintaining a compact 40~KB footprint, operating with only 3.25~MB of memory, and achieving 1.42~ms inference latency, our system meets the stringent efficiency demands of edge deployments. It achieves orders-of-magnitude faster processing than traditional ML pipelines and matches or exceeds the efficiency of specialized early-detection models while simultaneously delivering the highest accuracy, earliest detection, and lowest false detection rates across all evaluated datasets.

\subsection{Sensitivity Analysis of Confidence Threshold}
\label{sec:results:sensitivity}

\begin{figure}
    \centering
    \includegraphics[width=\textwidth]{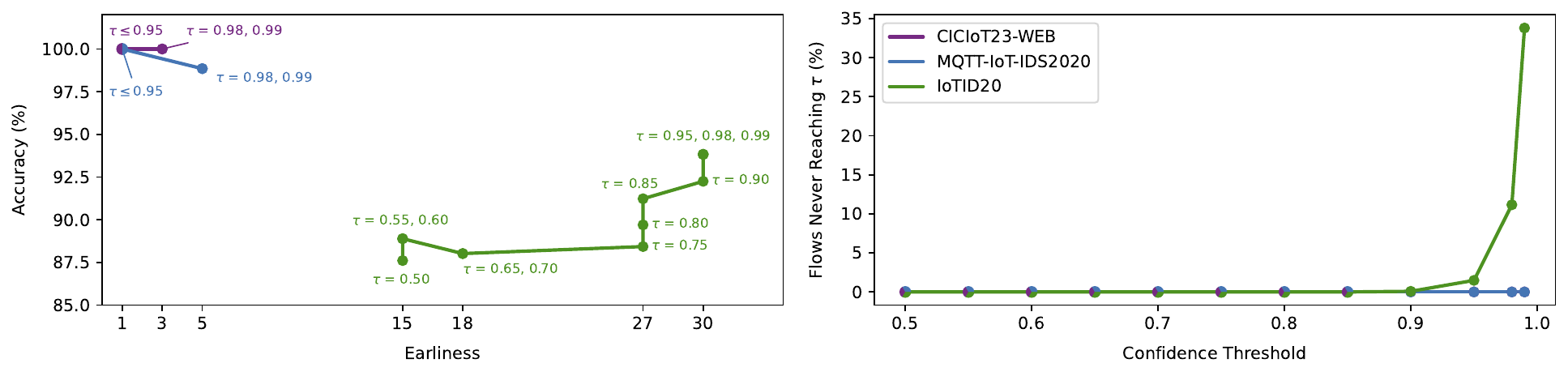}
    \caption{Sensitivity analysis of the confidence threshold ($\tau$). The left panel illustrates the trade-off between classification accuracy and earliness (in packets) as $\tau$ varies. The right panel displays the percentage of network flows that fail to reach the target confidence threshold within the maximum sequence length limit ($N=30$) for each $\tau$ value.}
    \Description{This figure presents a two-panel sensitivity analysis of the confidence threshold tau used for early decision-making in A-THENA. The left panel shows the relationship between classification accuracy and earliness for different values of tau across multiple datasets. Each curve corresponds to a dataset and consists of several points representing increasing values of tau. As tau increases, accuracy generally improves while earliness decreases, illustrating a trade-off between making earlier predictions and achieving higher confidence. For the IoTID20 dataset, this trend is clearly visible and mostly monotonic, with lower tau values yielding earlier but less accurate predictions, and higher values yielding more accurate but slightly delayed predictions. A small dip in accuracy is observed for intermediate values (0.65--0.75), indicating a transitional regime where predictions are neither sufficiently early nor sufficiently confident. The right panel shows the percentage of flows that never reach the specified confidence threshold for different values of tau. As tau approaches 1.0, this percentage increases, particularly for the IoTID20 dataset, indicating that overly strict thresholds may prevent the model from making a decision for some inputs. For moderate values of tau, the percentage remains close to zero across datasets, suggesting reliable decision coverage in practical operating ranges. Together, the two plots demonstrate that tau provides a controllable trade-off between detection speed, accuracy, and coverage, with stable behavior across a broad range of values.}
    \label{fig:confidence_threshold}
\end{figure}

To evaluate \athena’s flexibility, we analyze the impact of the confidence threshold $\tau$. Figure~\ref{fig:confidence_threshold} shows how varying $\tau \in [0.50, 0.99]$ affects the Accuracy--Earliness trade-off (left) and the fraction of flows that do not reach the threshold before reaching the maximum sequence length (right).

The left panel highlights the trade-off between detection speed and accuracy, most evident in the \iotid dataset. Lower thresholds (\textit{e.g.}, $\tau \le 0.60$) enable rapid decisions (within 15 packets) but limit accuracy to about 88.9\%. The relationship is not strictly monotonic: a slight accuracy dip appears for $\tau \in [0.65, 0.75]$, likely due to ambiguous intermediate packets. Higher thresholds ($\tau \ge 0.85$) force the model to wait for later packets, resolving ambiguity and achieving peak accuracy (93.83\% at $\tau = 0.95$). In contrast, \ciciot and \mqtt exhibit near-perfect accuracy with single-packet earliness for $\tau \le 0.95$, with only minor degradation at extreme thresholds. The right panel shows that overly high thresholds introduce instability. While $\tau \le 0.95$ is consistently reached, extreme values ($\tau \ge 0.98$) in \iotid lead to up to 33\% of flows failing to meet the threshold within 30 packets. These cases revert to full-sequence predictions, increasing delay and slightly reducing accuracy.

Overall, $\tau = 0.95$ provides the best balance, maximizing accuracy while avoiding instability. Lowering $\tau$ offers negligible practical benefit, as full-sequence processing latency is only 1.42 ms on a Raspberry Pi Zero 2 W (Section~\ref{dl_latency_mf}). Thus, prioritizing accuracy is preferable. Ultimately, $\tau$ serves as a highly adaptable hyperparameter. Practitioners can dynamically tune the threshold to match specific network constraints---lowering it for ultra-low-latency environments where microsecond-level reactions are critical, or maintaining it at strict levels to ensure near-zero false alarms in sensitive deployments.

\subsection{Case Studies: Visualizing Temporal Attention Dynamics}
\label{sec:case_studies}

\begin{figure*}[t]
    \centering
    \includegraphics[width=\linewidth]{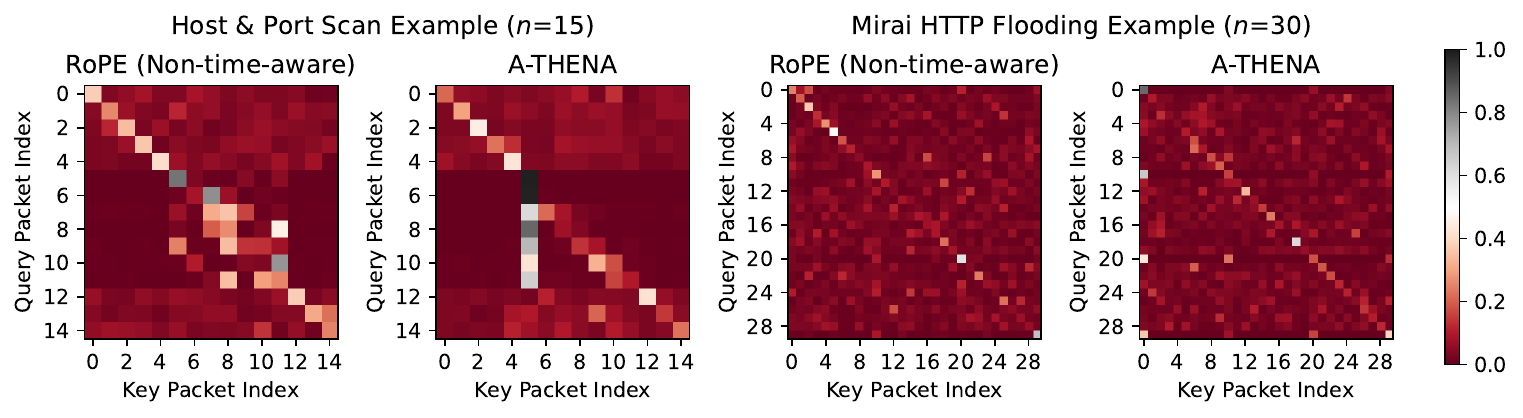}
    \caption{Attention weight heatmaps comparing \athena (TA RoPE) and RoPE on two representative flows from \iotid. (Left pair) Host \& Port Scan flow ($n=15$). (Right pair) Mirai HTTP Flooding flow ($n=30$).}
    \Description{Four side-by-side heatmaps displaying self-attention weights (Query Packet Index vs. Key Packet Index) for two attack scenarios from the IoTID20 dataset. The color scale ranges from dark red (0.0) to dark gray/black (1.0). The first two plots show a 15-packet Host and Port Scan. The RoPE model shows a smeared, blocky red/orange attention cluster between indices 5 and 11. The A-THENA model shows a sharp, dark vertical line of high attention firmly anchored at index 5. The last two plots show a 30-packet Mirai HTTP Flooding attack. The RoPE model shows a noisy, speckled heatmap with attention scattered across the sequence. The A-THENA model shows a distinct grid-like pattern with sharp, high-attention nodes isolated at periodic indices 0, 10, 20, and 29, cleanly filtering out the background noise.}
    \label{fig:attention_heatmaps}
\end{figure*}

To provide further insight into the behavior of the proposed Time-Aware Hybrid Encoding, we present two representative case studies from the \iotid dataset, visualizing the self-attention patterns learned by A-THENA and the respective non-time-aware baseline (RoPE). Figure~\ref{fig:attention_heatmaps} shows the attention weight matrices for two attack scenarios with different temporal characteristics. These case studies demonstrate that \athena leverages time-aware encodings to fundamentally reshape attention dynamics, enabling clearer separation of temporal patterns and more interpretable behavior.

\paragraph{Case Study 1: High-Frequency Bursts}
The left two panels of Figure~\ref{fig:attention_heatmaps} depict a 15-packet sequence from a \textit{Host \& Port Scan} attack. This attack is characterized by a rapid burst of reconnaissance packets (indices 5 through 11) sent with near-zero inter-arrival times. While both models recognize the malicious payloads within this window, their attention distributions differ significantly. RoPE produces a diffuse, block-like pattern over the burst, treating packets as merely sequential. In contrast, \athena exploits the near-zero inter-arrival times, sharply anchoring attention to the burst onset and forming a more structured representation of the attack.

\paragraph{Case Study 2: Periodic / Delayed Attacks}
The right two panels of Figure~\ref{fig:attention_heatmaps} illustrate a 30-packet sequence from a \textit{Mirai HTTP Flooding} attack, characterized by periodic malicious requests separated by deliberately long temporal gaps (\textit{e.g.}, indices 0, 10, 20, and 29). Under RoPE, attention is diluted across the full 30-packet sequence, yielding a near-uniform distribution; although the model captures "Keep-Alive" payloads, it cannot leverage timing gaps, resulting in noisy, unstructured attention. In contrast, \athena encodes these gaps as larger positional distances, causing these packets to emerge as temporal outliers. This produces a sharp grid-like pattern, where periodically delayed packets attend strongly to the initial connection and to each other, effectively capturing the attack’s temporal signature.

\subsection{Ablation Study: Dissecting \athena's Components}
\label{sec:results:ablation}

\begin{figure}
    \centering
    \includegraphics[width=\textwidth]{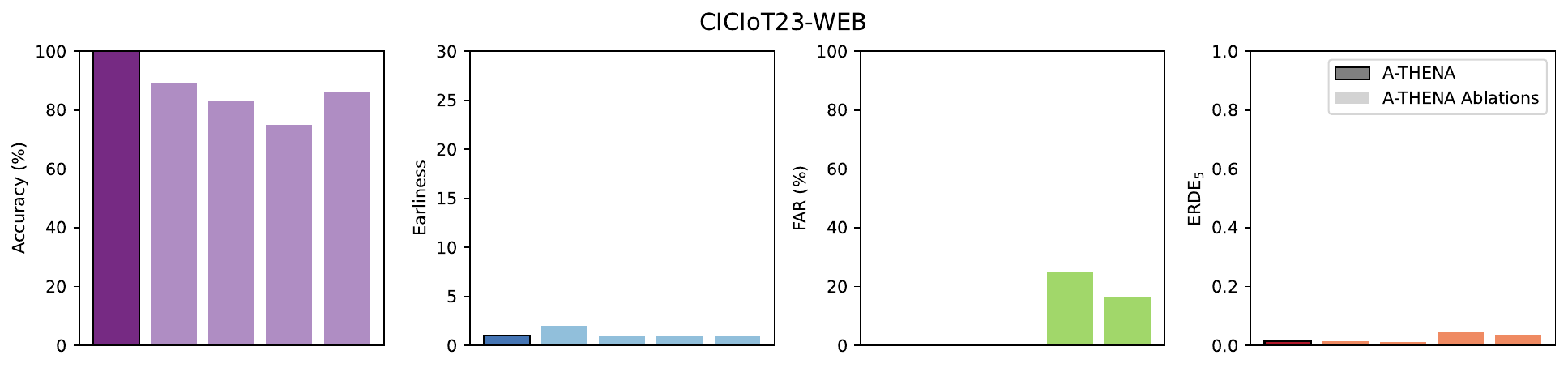} \\
    \includegraphics[width=\textwidth]{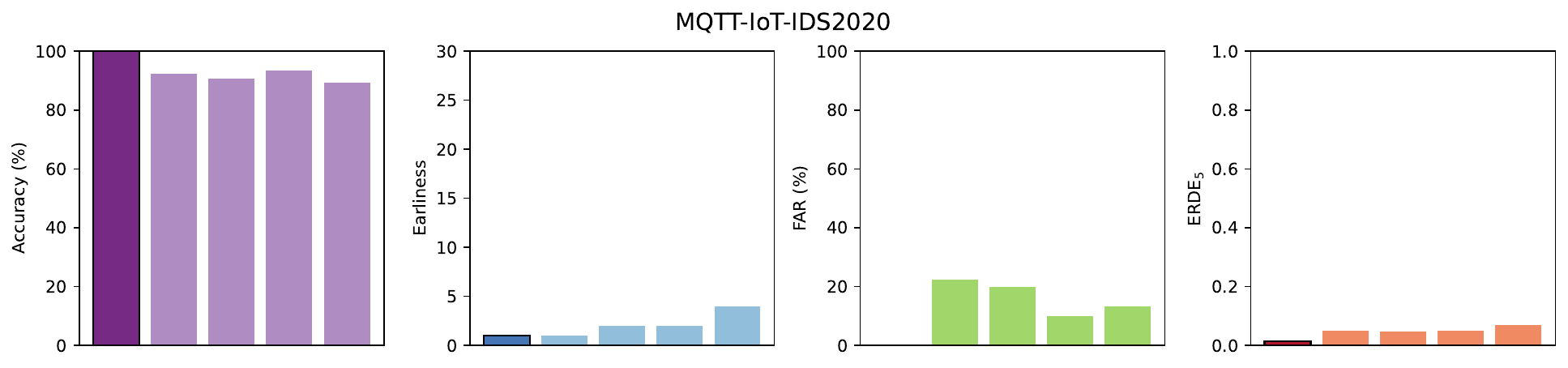} \\
    \includegraphics[width=\textwidth]{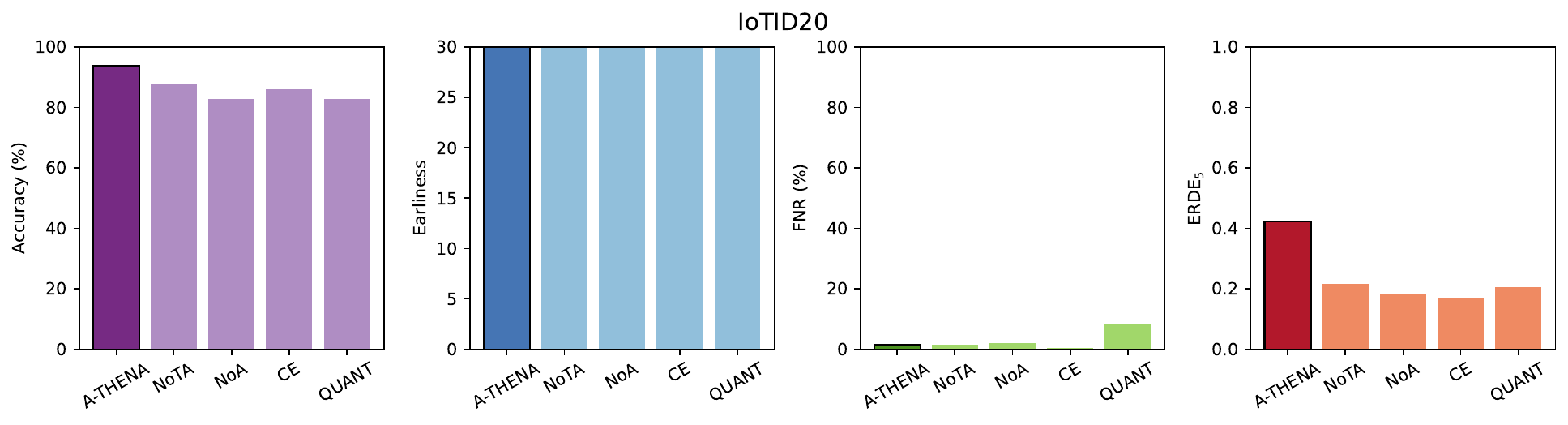}
    \caption{Impact of augmentation, EDL, and quantization on \athena's performance.}
    \Description{This figure compares the performance of the \athena system to its ablated variants across three IoT intrusion detection datasets: \ciciot (top row), \mqtt (middle row), and \iotid (bottom row). Each row contains four bar charts evaluating models based on Accuracy (\%), Earliness, False Alarm Rate (FAR or FNR), and Early Risk Detection Error at 5 packets (ERDE_5). The x-axis across all charts includes the original \athena and four ablation configurations: NoTA (without time-aware data augmentation), NoA (without data augmentation), CE (with cross-entropy loss instead of EDL), and QUANT (quantized version). \athena results are shown in solid dark bars, while ablations are in lighter shades. Across datasets, \athena generally achieves the highest accuracy and lowest error metrics, demonstrating the contribution of each component. For example, omitting time-aware augmentation or EDL slightly degrades accuracy and increases false alarms or ERDE rates. This figure highlights the effectiveness and resilience of \athena's design.}
    \label{fig:ablation}
\end{figure}

This subsection examines the core design components of \athena, including the online augmentation pipeline, the Early Detection Loss (EDL) function, and the impact of quantization. To analyze the role of augmentation, NoTA (No Timestamp Augmentation) applies only packet augmentation, omitting the two time-aware techniques (Jitter Injection and Traffic Scaling). In contrast, NoA (No Augmentation) removes all five augmentation methods, providing a baseline for assessing the impact of online data augmentation on early intrusion detection. The CE (Cross-Entropy) variant replaces \athena's EDL loss function with a traditional cross-entropy loss, allowing an evaluation of the effectiveness of loss-based optimization for early detection. Lastly, QUANT applies post-training INT8 quantization, measuring the trade-off between model compression and detection performance.

Figure~\ref{fig:ablation} summarizes the performance of \athena and its ablated variants across all three datasets, illustrating the contribution of each component to the overall system. The results strongly support the architectural choices underlying \athena, showing that every component materially enhances IoT intrusion detection effectiveness. In particular, incorporating time-aware augmentation yields substantial gains in accuracy and robustness, consistently outperforming packet-only augmentation. These findings highlight the importance of timestamp-aware transformations and specialized early-detection optimization for achieving reliable real-time intrusion detection in IoT settings.

\begin{table}
    \small
    \caption{Quantization Benefits}
    \label{tab:quant}
    \begin{tabular}{r | r r}
        \toprule
        \textbf{System} & \textbf{Latency Speedup} & \textbf{Memory Reduction} \\
        \midrule
        eRNN~\cite{ahmad24icstw}            & 0.86$\times$ & 1.04$\times$ \\
        eTransformer~\cite{islam23cloudcom} & 1.52$\times$ & 1.77$\times$ \\
        eAtt~\cite{ahmad23icstw}            & 1.43$\times$ & 1.08$\times$ \\
        eGlo~\cite{ahmad22icstw}            & 1.25$\times$ & 1.13$\times$ \\
        \midrule
        \textbf{\athena}                             & 1.37$\times$ & 1.00$\times$ \\
        \bottomrule
    \end{tabular}
\end{table}

\paragraph{Quantization Analysis}
Table~\ref{tab:quant} presents the effects of post-training INT8 quantization on both latency speedup and memory reduction for \athena and related work models. The results show that quantization mainly enhances latency performance for \athena, achieving a 1.37$\times$ improvement. However, no reduction in memory footprint is observed. This lack of memory reduction is likely due to the inherent memory overhead associated with the LiteRT runtime. Specifically, LiteRT's memory arena, which manages intermediate tensors during model inference, imposes a fixed memory overhead. This overhead can be substantial, sometimes even exceeding the model's own size. Consequently, irrespective of how compact the model is, this baseline memory requirement remains relatively constant, limiting potential memory savings through quantization. Among the evaluated architectures, the eTransformer model experiences the most significant benefits, achieving a latency speedup of 1.52$\times$ and a substantial memory reduction of 1.77$\times$. Other architectures demonstrate comparatively marginal improvements in these metrics.

The limited effectiveness of quantization for \athena can further be explained by its already compact Transformer architecture, which offers minimal scope for additional compression. Conversely, larger models like eTransformer, containing over one million parameters, substantially benefit from reducing high-precision weights to INT8, resulting in considerable memory and latency improvements. Additionally, the slight performance degradation observed in the eRNN model (0.86$\times$ latency speedup) implies that out-of-the-box post-training quantization may introduce inefficiencies in recurrent computations.

%% file: sections/7_conclusion.tex
This study introduced \athena, an early intrusion detection system that advances time-aware modeling for IoT security. By combining a hybrid positional encoding mechanism, a tailored augmentation strategy, and a custom loss function, the system captures fine-grained temporal signatures that characterize malicious behavior. Treating packet flows as non-uniform time series enables \athena to mitigate temporal aliasing effects and learn timing patterns that traditional encodings overlook. Experimental results across multiple IoT-focused datasets confirm its effectiveness, achieving near-perfect detection performance with minimal false alarms, while deployments on edge hardware validate its suitability for resource-constrained environments. Compared to prior work on time-aware encodings, \athena offers a unified hybrid framework, improved cross-dataset generalization, and new insights into early-stage intrusion detection across heterogeneous IoT protocols.

Although \athena demonstrates strong performance, several constraints remain that outline the boundaries of its current applicability. The approach depends on reliable packet timestamps, and environments with significant jitter or timestamp noise may require additional synchronization to ensure stable performance. Additionally, the adaptive flow aggregation strategy, while theoretically motivated, has not yet been stress-tested under extreme traffic loads, leaving its scalability beyond edge scenarios an open question. Furthermore, the system has not yet been evaluated against adversarial behaviors specifically crafted to manipulate temporal patterns, limiting guarantees on robustness. These constraints define natural directions for future research, including extending the framework to a broader spectrum of low-power platforms and increasingly hostile environments.